%% file: SN-9-11-1.tex
\documentclass[12pt]{article}
\usepackage{amsfonts,amssymb,amsmath,amscd,theorem,oldgerm,epsfig,units}
\usepackage[mathscr]{eucal}
\usepackage{mathrsfs}

\input diagram.tex 

\newtheorem{lemma}{Lemma}[section]
\newtheorem{proposition}[lemma]{Proposition}

\newtheorem{theorem}[lemma]{Theorem}

\newtheorem{definition}[lemma]{Definition}

\newtheorem{claim}[lemma]{Claim}

\newcommand{\EProof}{\hfill \blacksquare}

\def\ome {\omega}

\def\XI{\xi}

\def\lto{\longrightarrow}

\def\iso{\backsimeq}

\def\R{{\mathbb R}}
\def\Q{{\mathbb Q}}
\def\N{{\mathbb N}}
\def\Z{{\mathbb Z}}
\def\C{{\mathbb C}}
\def\Fq {\mathbb{F}_{q}}   

\def\Lm{\Lambda^*}

\def\GL{\mathrm{GL}}
\def\PGL{\mathrm{PGL}}

\def\G{\mathrm{G}}     

\def\AT{\mathbb{T}}      

\def\CA{{\mathrm T}_{\mathrm {A}}}

\def\Sp {\mathrm{Sp}} 
\def\E {\mathrm{H}}      
\def\Heis {\mathrm{H}}

\def\S0{S}

\def\T{{\mathrm{T}}}
\def\Td{\AT^\vee} 

\def\Irr{\mathrm{Irr}({\cal A} _\hbar)}
\def\A{\cal A}
\def\Ad{ {\cal A} _\hbar}

\def\h{ \hbar }

\def\H{{\cal H}}

\def\Hc {{\mathcal{H}_\V}} 
\def\V{\mathrm{V}} 
\def\W{{\mathrm W}}

\def\rhoh{\rho _{_\h}}

\def\Pih{\pi_{_\hbar}}

\def\X{{\mathrm X}}

\def\P1{\mathbb{P}^1}

\def\i_XI{i_{_{\XI}}}

\def\p_XI{p_{_{\XI}}}

\def\SE{\mathcal{E}}

\def\SF{\mathcal{F}}
\def\SG{\mathcal{G}}
\def\SL{\mathscr L}

\def\SK {\mathcal{K}} 

\def\Tr{\mathrm{Tr}}

\def\ev{{{\texttt{e.v.}}}}

\def\End{\mathrm{End}}
\def\Hom{\mathrm{Hom}}
\def\dim{\mathrm{dim}}

\def\Lag{\mathrm{Lag}}
\def\LGr{\mathrm{LGr}}

\def\L{\mathrm{L}} 
\def\M{\mathrm{M}} 
\def\N{\mathrm{N}} 

\def\Fr{\mathrm{Fr}}

\def\Qlb{\overline{\Q}_\ell}

\def\Perv{\mathrm{Perv}}
\def\w{\mathrm w}
\def\I{{\mathrm F}} 

\def\rev{\quad}

\def\r{\;}
\def\above{\overset}

\def\half{\begin{smallmatrix} \frac{1}{2} \end{smallmatrix}}

\def\n{\mathrm{n}}
\def\m{\mathrm{m}}

\begin{document}

\title{\large \texttt{HEISENBERG REALIZATIONS, EIGENFUNCTIONS AND PROOF OF THE KURLBERG-RUDNICK SUPREMUM CONJECTURE}}
\author{\small \textsf{SHAMGAR GUREVICH AND RONNY HADANI}\\\\ {\it Preliminary
version}}
\date{}
\maketitle
\bigskip

\numberwithin{equation}{subsection}

 \setcounter{section}{-1}

\begin{abstract}
In this paper, proof of the {\it Kurlberg-Rudnick supremum
conjecture} for the quantum Hannay-Berry model is presented. This
conjecture was stated in P. Kurlberg's lectures at Bologna 2001
and Tel-Aviv 2003. The proof is a primer application of a
fundamental solution: all the Hecke eigenfunctions of the quantum
system are constructed. The main tool in our construction is the
categorification of the compatible system of realizations of the
Heisenberg representation over a finite field. This enables us to
construct certain "perverse sheaves" that stands motivically prior
to the Hecke eigenfunctions.
\end{abstract}

\tableofcontents

\section{Introduction}
In most branches of modern science, systems are described by
mathematical equations. For example, in quantum mechanics one has
to consider the eigenfunction and eigenvalue problems of the
well-known \textit{Schr\"{o}dinger} equation:
\begin{equation} \label{schrodinger}
(-\frac{\hbar ^2}{2m} \Delta + \V) \Psi = \lambda \Psi,
\end{equation}
on some configuration space X. One of the main problems is to
obtain an effective description of the solutions of such
equations. This problem of-course becomes quite intractable as the
equation becomes more complicated.

\subsection{Quantum chaos}
A classical mechanical system is modelled by a phase space M. A
point of the space records the position and momenta of all the
particles in the system. The evolution of the system in time
defines a path $\gamma(t)$ in  M. Newton's laws say that this path
is the solution of the differential equation:
\begin{equation}\label{Newton}
\frac{d\gamma}{dt} = \xi_\mathrm{H}(\gamma(t)),
\end{equation}
where $\xi_\mathrm{H}$ is called the Hamiltonian vector field and
$\Heis$ is a function on M which corresponds to the energy of the
system. A physical observable (temperature, position,
momentum,...) corresponds to a function $f$ on M. If the system is
in a state $\gamma(t)$, then the outcome of the observation
corresponding to $f$ is a single value $f(\gamma(t))$.
\\
\\
A quantum mechanical system is modelled by an Hilbert space $\H$.
A vector $\Psi$ in $\H$ records the state of the system. The
evolution of the system in time therefore corresponds to a path
$\Psi(t)$ in $\H$. The physical laws of the system are encoded by
a self-adjoint operator $\Heis_\h $. The parameter $\h$ is called
the Planck constant. These laws state that if $\Psi(t)$ is a
vector that represents the system at time $t$, then it satisfies
an analogue of (\ref{Newton}):
\begin{equation}\label{timeSchrodinger}
\partial_t \Psi(t) = \Heis_\h \Psi(t).
\end{equation}
A solution $\Psi$ of equation (\ref{timeSchrodinger}) does not
represents a classical trajectory. Instead, it is interpreted as a
statistical entity in the following sense. In general, a physical
observable is an operator $\Pih(f)$ on $\H$. If the system is in
the state $\Psi$ then the observation that corresponds to
$\Pih(f)$ cannot be predicted exactly. There is a probability
distribution of possible outcomes which is determined by $\Psi$
and the average is given by the matrix coefficient
$\langle\Pih(f)\Psi,\Psi\rangle$.

Consider a chaotic classical dynamical system. The main objective
(cf. \cite{M, S}) of quantum chaos theory is to explain how
chaotic behavior is manifested at the quantum-mechanical level, or
at least in the semi-classical limit as $\h$ tends to $0$. During
the 70's and the 80's Berry, Berry-Tabor, Bohigas, Gianoni and
Schmidt obtained (cf. \cite{Be, H, M, S}) very accurate
conjectural descriptions of the behavior of eigenfunctions and
eigenvalues for generic chaotic systems, albeit not much has been
confirmed mathematically.
\subsection{The Hannay-Berry Model}
It is for the above reasons that the physicists  J. Hannay and
M.V. Berry proposed around 1980 \cite{HB} a simple mathematical
model for quantum mechanics on the two-dimensional torus $\mathbb
T$. In this model, one considers an ergodic (discrete) dynamical
system generated by a single linear map:
$$
\mathrm{A}:\mathbb{T}
\lto \mathbb{T}.
$$
The associated quantum system (cf. \cite{GH1, GH2, KR1}) is an
operator $\rhoh(\mathrm{A})$ acting on a finite dimensional
Hilbert space $\H$.

\subsection{The Kurlberg-Rudnick conjectures}
The Kurlberg-Rudnick conjectures is a set of conjectural
statements \cite{Ku1, Ku2, KR3, R1, R2}  describing the operator
$\rhoh(\mathrm{A})$. They were motivated by a series of
fundamental papers (cf. \cite{DGI, KR1, KR2, KR3}). These
conjectures describe the behavior of the common eigenstates $\Psi$
of the Hecke symmetries of the operator $\rhoh(\mathrm{A})$,
i.e.,:
\begin{equation}\label{HEF}
\rhoh(\mathrm{B})\Psi = \lambda(\mathrm{B})\Psi,
\end{equation}
for every symmetry $\mathrm{B}$.

\subsection{Supremum and value distribution
conjecture}\label{VDC} This is a conjecture \cite{Ku1, Ku2} on the
size and  on the value distribution of the Hecke eigenstates
$\Psi$, (\ref{HEF}). The conjecture says that for $\h$ of the form
$\h=\frac{1}{p}$, where $p$ is a prime number, the Hecke
eigenstates are uniformly bounded, i.e.,:
\begin{equation}\label{S}
\|\Psi\|_{_\infty}\leq 2
\end{equation}
for any Hecke eigenstate $\Psi$. Moreover, the conjecture claims
that the value distribution of each of these $\Psi$'s  (suitably
twisted \cite{K2}) behaves like the trace of a random matrix in
$\mathrm{SU}(2)$.

Special cases of this conjecture were obtained in \cite{KR2} using
results established in a work of N. Katz \cite{K2}. However, in
general only the bound:
$$
\|\Psi\|_{_\infty} \leq p^{3/8},
$$
was available \cite{KR2}. In the current work, proof of the
supremum conjecture (\ref{S}) is presented. Moreover, we show that
this bound holds true in any realization of the quantum system
(corresponds to measuring different observables).

\subsection*{Acknowledgments}
We thank our Ph.D. adviser J. Bernstein for his interest and
guidance in this project. We thank P. Kurlberg and Z. Rudnick who
discussed with us their papers and explained their results. We
would like to thank D. Kazhdan for sharing his thoughts about the
possible existence of canonical Hilbert spaces. We thank D. Kelmer
for interesting discussions. S.G. would like to thank M. Baruch
and the Technion for the kind support during the preparation of
this work. Finally, we would like to thank P. Deligne for letting
us use his ideas about the geometrization of the Weil
representation which appeared in a letter he wrote to D. Kazhdan
in 1982. Deligne's ideas were the starting point for the
categorification procedure that appears in the current work.

\section{The Hannay-Berry Model}
\subsection{Classical phase space}
Our classical phase space is the 2n-dimensional symplectic torus
$(\AT,\omega)$. It is equipped with an action of the group
$\Gamma\simeq \Sp(2\n,\Z)$ acting by linear symplectomorphisms.
\\
\\
In more detail, consider the torus $\AT := \W/\Lambda$, where $\W$
is a 2n-dimensional real vector space, i.e., $\W\simeq \R^{2\n}$,
and $\Lambda$ $\subset \W$ is a full rank lattice, i.e.,
$\Lambda\simeq \Z^{2n}$. The symplectic structure on $\AT$ is
obtained from a skew-symmetric bilinear form $\omega:\W\times
\W\rightarrow \R $. We require $\ome$ to be integral, i.e.,
$\omega(\Lambda \times \Lambda ) \subset \Z$ and normalized, i.e.,
Volume($\AT$) = 1.
Denote by $\Gamma\subset \Sp (\W,\omega)$ the subgroup of elements
preserving the lattice $\Lambda$. We have $\Gamma\simeq
\Sp(2\n,\Z)$. The group $\Gamma$ naturally acts as the group of
linear symplectomorphisms of $(\AT,\omega)$.
\subsection{Quantization}
Our quantum object will be a pair of compatible representations of
a quantum algebra $\Ad$ and the group $\Gamma$.
\\
\\
In more detail, consider the algebra $\A$ of trigonometric
polynomials on the torus, namely functions which can be presented
as a finite linear combination of characters. Note that the
algebra $\A$ has as a basis the lattice of characters $\Td :=
\Hom(\AT,\C^*)$.
\\
\\
Next, we want to define a family of quantum algebras. Consider the
dual lattice $\Lambda^{\ast}\subset \W^{\ast}$, defined by
$\Lambda ^{\ast}=\{\xi\in \W^{\ast}:\xi(\Lambda)\subset \Z \}$.
Note that we can identify $\Lambda ^*$ with the lattice of
characters on $\AT$ by the following map:
\begin{equation*}
  \xi \in \Lm \longmapsto e^{2 \pi i <\xi, \cdot>} \in \r \Td.
\end{equation*}
We construct a family of (star, $\ast-$) algebras $\Ad$, i.e., for
each $\h \in \R$ we consider the algebra generated over $\C$ by
the symbols $\{s(\xi):\xi\in\Lambda^{\ast}\}$ satisfying the
relations $s(\xi+\eta)=e^{\pi
i\hbar\omega(\xi,\eta)}s(\xi)s(\eta)$. The family $\Ad$ form a
one-parametric deformation of the commutative algebra $\A$, that
is, ${\A}_0= \A$. The parameter $\hbar$ is called the Planck
constant. The symbols $\{s(\xi):\xi\in\Lambda^{\ast}\}$ form a
basis of $\Ad$. This allows us to identify, as vector spaces, the
algebras $\Ad$ and $\A$ for every value of $\hbar$. We will often
identify the symbol $s(\xi)$ with the element $\xi$ itself, in
order to save notation. Following \cite{Ri}, we call the algebra
$\Ad$ the {\it Rieffel's} (algebraic) quantum torus.
\\
\\
The group $\Gamma$ acts by automorphism on the algebra $\Ad$ via
the formula $\mathrm{B} \cdot s(\xi)=s(\mathrm{B}\xi)$ for every
$\mathrm{B}\in\Gamma$. This induces an action of $\Gamma$ on the
category $\mathrm{Rep} (\Ad)$ of ($\ast-$) representation of
$\Ad$, and hence on the set $\Irr$ of isomorphism classes of
irreducible algebraic representations of $\Ad$. More concretely,
given a representation $\pi:\Ad\rightarrow \End(\H)$ and an
element $\mathrm{B}\in\Gamma$,  we define a new representation
$\pi^{\mathrm{B}}:\Ad\rightarrow \End(\H)$ by $\pi^{\mathrm{B}}(f)
:= \pi(f \circ \mathrm{B})$.
\\
\\
For the remainder of this section, we fix $\hbar=\frac{1}{p}$,
where $p$ is an odd prime number. We have the following basic
theorem:

\begin{theorem}[\cite{GH2}]\label{equivariantquant} There exists a unique
$($up to isomorphism$)$ irreducible representation $(\pi,\Ad,\H)$
so that its isomorphism class $[\pi] \in \Irr$ is fixed by
$\Gamma$.
\end{theorem}

Let $(\pi,\Ad,\H)\in \Irr$, be a representation such that
$\mathrm{B}\cdot[\pi]=[\pi]$ for all $\mathrm{B} \in \Gamma$. This
is equivalent to having a projective representation
$\rho:\Gamma\rightarrow \PGL(\H)$ and
compatibility condition:%
\begin{equation}
\rho(\mathrm{B})\pi(f)\rho(\mathrm{B})^{-1}=\pi(f \circ
\mathrm{B}),\label{egorov}
\end{equation}

for every $\mathrm{B}\in\Gamma$ and $f\in \Ad$. Condition
(\ref{egorov}) is called, traditionally, the \textit{Egorov}
identity.

In fact one can do even better. The projective representation
$\rho$ can be linearized. There exists a canonical linearization
$\rho :\Gamma\rightarrow \GL(\H)$, which factors through the
finite quotient group $\G\simeq \Sp(2\n,\mathbb{F}_{p})$.
Altogether we can take our quantum object to
consist of a pair:%
\begin{eqnarray*}
\pi :\Ad &\rightarrow & \End(\H),
\\
\rho :\G & \rightarrow & \GL(\H),
\end{eqnarray*}
satisfying the Egorov identity (\ref{egorov}).
\\
\\
\textbf{Comments.}
\begin{enumerate}
\item All irreducible representations $\pi\in \Irr$ are of similar
nature. They are all finite dimensional, more precisely
$p$-dimensional. The set of equivalence classes of irreducible
representations is a manifold, which is a principal homogeneous
space over the torus $\AT$ (cf. \cite{GH2}).

\item The representation $\rho:\G\rightarrow \GL(\H)$ is the celebrated Weil
representation \cite{W1} of the finite symplectic group. This
representation is obtained here via quantization of the torus.
This approach is different from the classical constructions
\cite{Ge} and extend an earlier work carried out by the authors in
the two-dimensional setting \cite{GH3}.
\end{enumerate}

\subsection{Classical and quantum dynamical systems}

Here we restrict our attention to the case $\n=2$, namely we deal
with the two-dimensional symplectic torus $(\AT,\omega)$.
\\
\\
\textbf{Classical system.} We fix an hyperbolic element
$\mathrm{A}\in\Gamma$, namely, an element for which all
eigenvalues in $\overline{\Q}$ are not roots of unity. Consider
the corresponding automorphism
\begin{equation}\label{A-action}
\mathrm{A}: \AT \lto \AT.
\end{equation}
It is well-known that, via the iterations of the action
(\ref{A-action}), the element $\mathrm{A}$ generates a discrete
ergodic dynamical system on the torus $\AT$.
\\
\\
\textbf{Quantum system.} Taking the element $\mathrm{A}$, now
considered as an element of the finite group $\G$, we obtain a
unitary operator:
\begin{equation}\label{quantum-action}
\rho(\mathrm{A}):\H\rightarrow \H.
\end{equation}
The operator (\ref{quantum-action}) is considered to be the
\textit{quantization} of the classical dynamical system
$\mathrm{A}:\AT\rightarrow \AT$.

\section{Universal Supremum Bound Conjecture\label{problem}}
It is the main meta-question in the area of quantum chaos to
investigate: What manifestations of the chaotic behavior of
$\mathrm{A}$ are seen at the quantum mechanical level? In what
follows we are going to study a specific aspect of this question
\cite{Be, H, KR3}.

\subsection{Hecke eigenvectors}
Denote by $<\mathrm{A}>\subset \G$ the cyclic group generated by
the element $\mathrm{A}$. It is contained inside a "slightly"
bigger group $\T_{\mathrm{A}}$, namely, the centralizer of the
element $\mathrm{A}$ in $\G$. The group $\T_{\mathrm{A}}$ is an
algebraic group, more precisely it consists of the rational points
of an algebraic torus. We follow \cite{KR1} and call
$\T_\mathrm{A}$ the \textit{Hecke} torus.

The torus $\T_{\mathrm{A}}$ acts on $\H$ in a semi-simple fashion.
Hence, we obtain a decomposition into character spaces:
\[
\H=%
{\textstyle\bigoplus\limits_{\chi:\T_{\mathrm{A}}\rightarrow
\C^{\ast}}}
\H_{\chi}.%
\]

The group $\T_{\mathrm{A}}$ acts on the subspace $\H_{\chi}$ by
the character $\chi$. The subspace $\H_{\chi}$ is called the
$\chi$-Hecke eigenspace, and a unit vector $v_{\chi}\in
\H_{\chi}$, $\left\Vert v_{\chi}\right\Vert =1$, is called a
(normalized) $\chi$-Hecke eigenvector.

\subsubsection{Universal supremum bound}

Fix a multiplicative character $\chi:\T_\mathrm{A} \rightarrow
\C^{\ast}$, and let $v_{\chi}\in \H_{\chi}$ be a (normalized)
Hecke eigenvector.
The elements $\xi\in\Lambda^{\ast}$, gives a standard generating
set of observables $\{\pi(\xi)\}_{\xi\in\Lambda^{\ast}}$. Choosing
an observable $\pi(\xi)$, we can realize $\H\simeq \H_{\xi} :=
\Gamma(\sigma(\xi),\mathcal{\H}_{\xi})$, i.e. the global section
of an Hermitian line bundle $\mathcal{\H}_{\xi}$, on the spectrum
$\sigma(\xi)$ of the operator $\pi(\xi)$. This line bundle is
described by having fibers $(\mathcal{\H}_{\xi})_{x} =\H_{x}$ for
every $x\in\sigma(\xi)$ on which we have the action
$\pi(\xi)_{|\H_{x}}=x\cdot \mathrm{Id}.$ The vector $v_{\chi}$
gives a global section $\Psi_\chi \in \H_{\xi}$. We denote by
$\parallel \Psi_\chi\parallel_{_\infty}$ its supremum:
$$
\parallel \Psi_\chi\parallel_{_\infty} := \underset{x\in\sigma(\xi)}{\mathrm{Sup}}\left\vert \Psi_\chi(x)\right\vert.
$$

In this paper we are going to prove the following extended version
of the {\it Kurlberg-Rudnick supremum conjecture} (cf. \cite{Ku1,
Ku2, KR3}):

\begin{theorem}[Universal supremum bound]\label{universal}
We have:
$$
\parallel \Psi_\chi\parallel_{_\infty} \, \leq 2.
$$
Note, that the bound is independent of the character $\chi$, the
observable $\xi$ and the Planck constant $\ \hbar=\frac{1}{p}$.
\end{theorem}
\textbf{Comments.}
\begin{enumerate}
\item \textbf{Physical interpretation.} The unit sphere $S(\H)=\{v\in \H:$
$\left\Vert v\right\Vert =1\}$ constitutes the set of (pure)
quantum states. Measuring a quantum observable $\pi(\xi)$ on a
quantum state $v\in S(\H)$, amounts to realizing $v$ as a section
$\Psi_{v}\in\mathcal{\H}_{\xi}.$ The possible outcomes of the
measurement procedure are elements $x\in\sigma(\xi)$, with
probabilities $\left\vert \Psi_{v}(x)\right\vert ^{2}$. The
statement of Theorem \ref{universal} may be interpreted as saying
that an Hecke eigenvector $v_{\chi}$ is a special quantum state,
on which measurement of any standard observable gives outcomes in
fairly uniform distribution.

\item \textbf{Mathematical interpretation.} Suppose we are looking at a specific vector
$v\in \H$ in two different realizations $\H_{\xi}$ and $\H_{\eta}$
corresponding to the observables $\xi,\eta\in\Lambda^{\ast}$, that
are assumed to be non-proportional. The associated sections
$\Psi_v^{\xi}\in \H_{\xi}$ and $\Psi_v^{\eta}\in \H_{\eta}$ are
related by a certain kind of Fourier transform. Hence, Theorem
\ref{universal} may be interpreted as saying that $\Psi_v$
satisfies $\left\vert \Psi_v(x)\right\vert \leq 2,\,$and also
$\left\vert \widehat{\Psi}_v(x)\right\vert \leq 2,$ where
$\widehat{\Psi}_v$ is the Fourier transform of $\Psi_v$. This is
of-course not a trivial property.

\item \textbf{Implication of the Kurlberg-Rudnick supremum conjecture.}
Fix a prime $p$, and choose a standard observable $\xi$. The
realization $(\rho,\G,\H_{\xi})$ amounts to the standard
realization of the Weil representation on the space of functions
$\mathrm{L}^{2}(\mathbb{F}_{p},\C)$. Consider an Hecke eigenvector
$\Psi_\chi\in$ $\mathrm{L}^2(\mathbb{F}_{p},\C)$. The
Kurlberg-Rudnick conjecture states that  $|\Psi_\chi(x)|\leq 2$
for every $x\in\mathbb{F}_{p}$. Note, that this bound is
independent of the prime $p$. This is of-course a particular case
of Theorem \ref{universal}. The element $\mathrm{A}$ being
hyperbolic implies that for half of the primes $p$, the torus
$\T_{\mathrm{A}}$ is inert (does not split). We want to mention
here that the main problem is to compute $\Psi_\chi$ for $p$ where
$\T_{\mathrm{A}}$ is inert.
\end{enumerate}

\section{Canonical Hilbert Space}

Let $(\V,\omega)$ be a 2n-dimensional symplectic vector space over
the finite
field $\mathbb{F}_{q}$. Fix a non-trivial additive character $\psi:\mathbb{F}%
_{q}\rightarrow \C^{\ast}$. We have the following theorem (cf.
\cite{GH3}):

\begin{theorem}[Canonical Hilbert space]\label{canonicalhilbert}

There exists a canonical Hilbert space $\Hc$ associated to the
pair $(\V,\psi)$.
\end{theorem}

An immediate consequence of this theorem is that all symmetries of
$(\V,\omega)$ automatically act on $\H_{\V}$. In particular, we
obtain a (linear) representation of the group $\G
:=\Sp(\V,\omega)$ of linear symplectomorphisms.
\subsection{Construction}

\subsubsection{Heisenberg group}

There exists a two-step nilpotent group $\E=\E(\V,\ome)$, called
the Heisenberg group, associated to the symplectic vector space
$(\V,\omega)$. As a set $\E =\V\times\mathbb{F}_{q}$, with the
following multiplication rule:
$$
(v,z)\cdot(v^{\prime},z^{\prime })=(v+v^{\prime},z+z^{\prime}+
\half \omega(v,v^{\prime})).
$$
The center of $\E$ is $Z(\E)=\{(0,z):z \in\mathbb{F}_{q}\}$.
Identifying $Z(\E)=\mathbb{F}_{q}$, we consider the character
$\psi$ to be a character of the center $Z(\E)$. We have the
following fundamental theorem:

\begin{theorem}[Stone-Von Neumann]

There exists a unique $($up to isomorphism$)$ irreducible
representation $(\pi,\E,\H)$ with central character $\psi$, i.e.,
$\pi_{|Z(\E)} =\psi$.
\end{theorem}

The representation $(\pi,\E,\H)$ is called the \textit{Heisenberg
}representation.

\subsubsection{Models\label{models}}
Although the Heisenberg representation is unique, it admits a
multitude of different models (realizations). Here we construct a
specific family of such models. A fundamental ingredient in our
construction is the notion of \textit{enhanced
Lagrangian}\footnote{We thank A. Polishchuk for pointing out to us
that this should be thought of as an $\Fq$-analogue of well-known
considerations with usual oriented Lagrangians giving explicitly
the metaplectic covering of $\mathrm{Sp}(2\n,\R)$ (cf.
\cite{LV}).} suggested to us by J. Bernstein \cite{B, GH3, GH4}.
Consider the set $\Lag := \LGr(\V,\omega)$ of all Lagrangian
subspaces in $\V$. The set $\Lag$ is called the {\it Lagrangian
Grassmannian} associated to $\V$.

\begin{definition}
An enhanced Lagrangian is a pair $(\L,\sigma_{\L})$, where $\L\in
\Lag$, and $0\neq\sigma_{\L}\in\Lambda^{\n}\L$.
\end{definition}

The set of enhanced Lagrangians is denoted by $\Lag^{\circ}$.
Consider an element $\L^{\circ}\in \Lag^{\circ}$,
$\L^{\circ}=(\L,\sigma_{\L})$. Let
$\widetilde{\L}=pr^{-1}(\L)=\L\times Z(\E)$, where
$pr:\E\rightarrow \V$ is the standard projection. The set
$\widetilde{\L}\subset \E$ is an abelian subgroup of $\E$. \
Define the character $\psi_{\L^{\circ}}:\widetilde{\L}\rightarrow
\C ^{\ast}$ \ as an extension of $\psi$: $\psi_{\L^{\circ}}(l,z
)=\psi(z)$. Associated to this data we construct an Hilbert space
$\H_{\L^{\circ}}=\H(\E,\L^{\circ},\psi)=\{f:\E\rightarrow \C
:f(\widetilde{l}\cdot h)=\psi_{\L^{\circ}}(\widetilde{l})f(h)\}$.
This is a particular case of induction, equivalently we can write
$\H_{\L^{\circ}}=\mathrm{Ind}_{\widetilde{\L}}^{\E}(\psi_{\L^{\circ}})$.
The Hilbert space $\H_{\L^{\circ}}$ admits a representation of
$\E$, acting via right multiplication. We denote this
representation by $(\pi_{_{\L^{\circ}}},\E,\H_{\L^{\circ}})$. It
is not hard to see that $\pi_{_{\L^{\circ}}}$ is an irreducible
representation, with central character $\psi$, thus it constitutes
a model of the Heisenberg representation.

\subsubsection{Canonical intertwining operators}

Given a pair $(\L^{\circ},\M^{\circ})\in \Lag^{\circ}\times
\Lag^{\circ}$, we have
two models $(\pi_{_{\L^{\circ}}},\E,\H_{\L^{\circ}})$ and $(\pi_{_{\M^{\circ}%
}},\E,\H_{\M^{\circ}})$. Consider the space
$\mathrm{Int}_{\M^{\circ},\L^{\circ}}$ of intertwining operators:
$$
\mathrm{Int}_{\M^{\circ},\L^{\circ}}:=\Hom_{\E}(\H_{\L^{\circ}%
},\H_{\M^{\circ}}).
$$
Both models are irreducible representations of $\E$ so we have
$\dim\,\mathrm{Int}_{\M^{\circ},\L^{\circ}}=1$. Every element
$\mathrm{F}\in \mathrm{Int}_{\M^{\circ},\L^{\circ}}$ is
proportional to an averaging operator:
\[
\mathrm{F}(f)(h)= \textsf{g}_{_\mathrm{F}}%
{\textstyle\sum\limits_{m\in \M}}
f(m\cdot h),
\]
for $f\in \H_{\L^{\circ}}$ and $h\in\E$. Here
$\textsf{g}_{_\mathrm{F}}$ is the proportionality coefficient. It
turns out that one can choose the elements $\mathrm{F}_{\M^{\circ
},\L^{\circ}}$ in a canonical fashion, and we shell discuss this
next.

Let $\G_{m}$ denote the (finite) multiplicative group, $\G_{m}=\mathbb{F}%
_{q}^{\ast}$. Given a Lagrangian $\L\in \Lag$, we have the group
$\G_{m}$ acting on the 1-dimensional vector space $\Lambda^{\n}\L$
by homoteties. This induces an action of $\G_{m}$ on the space
$\Lag^{\circ}$. It is given by the formula
$a\cdot(\L,\sigma_{\L})=(\L,a\cdot$ $\sigma_{\L})$ for $a\in
\G_{m}$. Denote by $\chi_{_\mathrm{q}}:\G_{m}\rightarrow
\C^{\ast}$ the Legendre quadratic character. We have the following
theorem:

\begin{theorem}[Canonical intertwining operators]\label{CIO}

There exists a canonical family
$\{\mathrm{F}_{\M^{\circ},\L^{\circ}}\in
\mathrm{Int}_{\M^{\circ},\L^{\circ}}\}$ characterized by the
following properties:

\begin{enumerate}
\item \textbf{Normalization.} $\mathrm{F}_{\L^{\circ},\L^{\circ}}=1.$

\item \textbf{Invariance.} $\mathrm{F}_{\M^{\circ},\L^{\circ}}^{g}=\mathrm{F}
_{g\M^{\circ},g\L^{\circ}}$ for every element $g\in \G$.

\item \textbf{Convolution.} $\mathrm{F}_{\N^{\circ},\M^{\circ}}\circ
\mathrm{F}_{\M^{\circ},\L^{\circ}}=\mathrm{F}_{\N^{\circ},\L^{\circ}}.$

\item \textbf{Sign rule.} $\mathrm{F}_{a\M^{\circ},\L^{\circ}}=\chi_{_\mathrm{q}}
(a)\mathrm{F}_{\M^{\circ},\L^{\circ}}$, and $\mathrm{F}_{\M^{\circ},a\L^{\circ}}%
=\chi_{_\mathrm{q}}(a)\mathrm{F}_{\M^{\circ},\L^{\circ}}$ for
every $a\in \G_{m}$.
\end{enumerate}
\end{theorem}

\textbf{Comment.} We elaborate on the formal meaning of Property 2
in Theorem \ref{CIO}. The group $\G$ acts on all structures
involved. The action on the space of enhanced Lagrangians
$\Lag^{\circ}$ is tautological. Also $\G$ acts on the Heisenberg
group by automorphism. This action is the standard action on the
vector space $\V$, and it is trivial on the center $Z(\E)$. This
induces an action of $\G$ on the space of functions
$\mathrm{\L}^2(\E,\C)$, given by $g\cdot f(h)=f(g^{-1}h)$. It is
easy to verify that $g$ sends the space $\H_{\L^{\circ}}$
isomorphically to the space $\H_{g\L^{\circ}}$. The notation
\textbf{\ }$\mathrm{F}_{\M^{\circ},\L^{\circ}}^{g}$ stands for the
composition $g\circ$\textbf{\
}$\mathrm{F}_{\M^{\circ},\L^{\circ}}\circ g^{-1}$. Now the
interpretation of Property 2 is clear.
\subsubsection{Canonical Hilbert space}
The canonical Hilbert space $\Hc$ is defined as the subspace
$\H_{\V} \subset {\textstyle\bigoplus\limits_{\mathbf{\
}L^{\circ}\in \Lag^{\circ}}} \H_{\L^{\circ}}$, with:
$$\Hc := \{(v_{\L^{\circ}})_{\L^{\circ}\in \Lag^{\circ}}:\,
v_{\M^{\circ}}=\mathrm{F}_{\M^{\circ},\L^{\circ}}(v_{\L^{\circ}})\r
\text{ for every pair}\r (\M^{\circ},\L^{\circ})\in
\Lag^{\circ}\times \Lag^{\circ}\}.
$$

In other words, the space $\H_{\V}$ consists of compatible systems
of vectors. The existence of such systems is a consequence of the
convolution property satisfied by the canonical intertwining
operators (see Property 3 in Theorem \ref{CIO})$.$

\subsection{Weil representation}

The vector space $\H_{\V}$, being canonical, admits automatically
a representation of the group $\G$, called the Weil
representation. We give here two explicit descriptions of the Weil
representation:

\textbf{(I) Invariant form.} Consider the map
$\rho_{_\V}:\G\rightarrow \GL(\H_{\V})$, which is defined by the
action on a compatible system
$\overrightarrow{v}=(v_{\L^{\circ}})_{\L^{\circ}\in
\Lag^{\circ}}$:
\begin{equation}
[\rho_{_\V}(g)(\overrightarrow{v})]_{\L^{\circ}}:=v_{g^{-1}\L^{\circ}}^{g},%
\label{Weilrep}%
\end{equation}
where the function $v_{g^{-1} \L^{\circ}}^{g}\in \H_{\L^{\circ}}$
is defined by the formula:
$$v_{g^{-1}\L^{\circ}} ^{g}(h)
 := v_{g^{-1}\L^{\circ}}(g^{-1}h),$$
for every $h\in \E$. It is a direct consequence of the properties
1-3 in Theorem \ref{CIO} that formula (\ref{Weilrep}) gives a
representation of $\G$.

\textbf{(II) Models.} Choosing an enhanced Lagrangian
$\L^{\circ}\in \Lag^{\circ }$, we can identify
$\H_{\L^{\circ}}\simeq \H_{\V}$ via the inclusion
$\H_{\L^{\circ}}\hookrightarrow%
{\textstyle\bigoplus\limits_{\M^{\circ}\in \Lag^{\circ}}}
\H_{\M^{\circ}}$. Under this identification, the representation
$(\rho
_{_\V},\G,\H_{\V})$ \ is realized as a representation $(\rho_{_{\L^{\circ}}%
},\G,\H_{\L^{\circ}})$, which is given by:
$$\rho_{_{\L^{\circ}}}(g)(v_{\L^{\circ}%
})=\mathrm{F}_{\L^{\circ},g\L^{\circ}}(v_{\L^{\circ}}^{g}).$$

\subsection{Back to quantum mechanics}
We would like to establish a dictionary between quantum mechanics
on the two-dimensional torus $(\T,\omega)$ (see Section
\ref{problem}), and the formalism of the canonical Hilbert space.
This type of equivalence is valid when the planck constant $\hbar$
takes rational values - rational quantization. We will be even
more particular, assuming that $\hbar=\frac{1}{p},$ where $p$ is
an odd prime.
\\
\\
Consider the lattice $\Lambda^{\ast} = \Hom(\T,\C^*)$ of
characters of $\T$. Let $\V=\Lambda^{\ast }/p\Lambda^{\ast}$ be
the quotient abelian group. The group $\V$ has a natural structure
of a two-dimensional vector space over the finite field
$\mathbb{F}_{p}.$ The form $\omega$ induces a symplectic form on
$\V$, which we denote also by $\omega:\V\times
\V\rightarrow\mathbb{F}_{p}$. In the definition of the algebra
$\Ad$, for $\hbar=\frac{1}{p}$, a character $e^{\frac{2\pi i
x}{p}}$ appeared. We denote it by $\psi,$ and consider it as a
character of the finite field $\mathbb{F}_{p}$.
\\
\\
The dictionary works as follows:

\begin{itemize}
\item The algebra $\Ad$ \ is replaced by the Heisenberg group $\E=
\E(\V,\ome)$.

\item The representation $(\pi,\Ad,\H)$ (see Theorem
\ref{equivariantquant}) corresponds to the Heisenberg
representation, realized on the canonical Hilbert space
$(\pi_{_\V},\E,\H_{\V})$.

\item The Weil representation $(\rho,\G,\H)$ is taken in its invariant form on
the canonical Hilbert space $(\rho_{_\V},\G,\H_{\V})$.

\item Choosing a standard observable $\xi\in$ $\Lambda^{\ast}$, and
identifying $\H\simeq \H_{\xi}$ corresponds to considering the
enhanced Lagrangian $\L_{\xi}^{\circ}=$ $(\L_{\xi} ,\sigma_{\xi})
:= (\mathbb{F}_{p}\cdot\xi,\xi)$ and identifying $\H_{\V}\simeq
\H_{\L_{\xi}}$. The vector space $\H_{\L_{\xi}}$
is obtained via induction so it is the space $\Gamma(\X_{\L_\xi^{\circ}}%
,\mathcal{\H}_{\L_{\xi}^{\circ}})$ of global sections of an
Hermitian line bundle $\mathcal{\H}_{\L_{\xi}^{\circ}}$ on the set
$\X_{\L_\xi^{\circ}}=\widetilde {\L_\xi}\backslash \E$. An Hecke
eigenvector $v_{\chi}\in \H_{\V}$ \ is realized as a global
section $\Psi_\chi \in \H_{\L_{\xi}^{\circ}}.$
\\
\\
Having the above dictionary, Theorem \ref{universal} can be
equivalently formulated as follows:
\end{itemize}

\begin{theorem}[Universal supremum bound - reformulation]
We have
$$\underset{x\in \X_{\L^{\circ}}}{\mathrm{Sup}}\left\vert \Psi_\chi(x)\right\vert \leq 2,$$
for every character $\chi$, every enhanced Lagrangian $\L^{\circ}$
and every prime $p$.
\end{theorem}

An interesting question is whether one can compute in some
effective manner the Hecke eigenvector $v_{\chi}$. We will
elaborate on this issue. Assume first that there exists an
enhanced Lagrangian $\L^{\circ}\in \Lag^{\circ}$,
$\L^{\circ}=(\L,\sigma)$, such that $\L$ is fixed by the Hecke
torus $\T_{\mathrm{A}}$, that is, $g\L=\L$ for every $g\in
\T_{\mathrm{A}}$. In this case, one can compute the Hecke
eigenvector $v_{\chi}$ in the realization $\H_{\L^{\circ}}$. In
more detail, let $\L^{\circ\prime}=(\L^{\prime},\sigma^{\prime})$
be an enhanced Lagrangian such that $\V=\L\oplus \L^{\prime}$, and
$\L^{\prime}$ is also fixed by $\T_{\mathrm{A}}$ (such a choice
always exists). Using $\L^{\prime}$ as a cross section
$s:\X_{\L^{\circ}}\rightarrow \E$ \ we identify
$\H_{\L^{\circ}}=\L^{2}(\L^{\prime},\C)$. It is easy to deduce
that in this case the torus $\T_\mathrm{A}$ splits. Moreover, it
is possible to choose the identification $\T_{\mathrm{A}}\simeq
\G_{m}$ so that the restriction of the Weil representation to
$\G_{m}$ acting in the realization $\L^{2}(\L^{\prime},\C)$ is
given by the following formula:
\[
[\rho_{_{\L^{\circ}}}(a)f](l^{\prime})=\chi_{_\mathrm{q}}(a)f(al^{\prime}).
\]

Now, identifying $\L^{\prime}\simeq\mathbb{F}_{p}$, by $x\mapsto
x\sigma ^{\prime}$, we take the function $\Psi_\chi \in
\L^{2}(\mathbb{F}_{p},\C)$, with:
\begin{equation}\label{e.v.}
\Psi_\chi(x):=\chi_{_\mathrm{q}}(x)\chi(x).
\end{equation}
It is clear that formula (\ref{e.v.}) describes a
$\chi$-eigenvector for $\CA$.
\\\\
\textbf{Summary.} We found a realization $\H_{\L^{\circ }}$ in
which the Hecke torus $\T_{\mathrm{A}}$ acts in a geometric
manner. In this realization we are able to compute precisely the
eigenvector $v_{\chi}$.

\textbf{Problem.} What to do when the Hecke torus is inert, i.e.,
does not split? In this case there exists no Lagrangian $\L\in
\Lag$, which is fixed by $\T_{\mathrm{A}}$. Hence, we will
approach the problem from a more abstract perspective.

\section{Geometrization}
Geometrization is a general methodology, invented by Grothendieck,
by which sets are replaced by algebraic varieties (over the finite field $\mathbb{F}_{q}%
$) and functions are replaced by sheaf theoretic objects
($\ell$-adic Weil sheaves). In this section we are going to apply
this methodology to obtain a geometric analogue of the canonical
Hilbert space\footnote{This is a generalization of the
geometrization of the Weil representation proposed by Deligne in
\cite{D1} and presented by the authors in \cite{GH3}.}. In
particular, we obtain the following: the canonical intertwining
operators (CIO) $\{\mathrm{F}_{\M^{\circ },\L^{\circ}}\}$ are
replaced by a single (shifted) perverse Weil sheaf. Each model
$(\rho_{_{\L^{\circ}}
},\G,\pi_{_{\L^{\circ}}},\E,\H_{\L^{\circ}})$ is replaced by a
category $\mathcal{D}_{\L^{\circ}}$ of Weil sheaves, equipped with
a compatible action of the groups $\E$ and $\G$. Moreover, there
is a formal procedure\footnote{Similar to, and very much
influenced by the procedure that appears in the work of
Braverman-Polishchuk \cite{BP}.} to reconstruct the representation
$(\rho_{_{\L^{\circ}}
},\G,\pi_{_{\L^{\circ}}},\E,\H_{\L^{\circ}})$ from the category
$\mathcal{D}_{\L^{\circ}}$.

\subsection{Functorial description of the CIO}

For every pair $(\M^{\circ},\L^{\circ})\in \Lag^{\circ}\times
\Lag^{\circ}$, the intertwining operator
$\mathrm{F}_{\M^{\circ},\L^{\circ}}:\H_{\L^{\circ}}\rightarrow
\H_{\M^{\circ}}$ is given by a kernel function
$\I_{\M^{\circ},\L^{\circ}}:\E\times \E\rightarrow \C$ satisfying
the following properties:

\begin{enumerate}
\item \textbf{Intertwining.}
\begin{eqnarray*}
\I_{\M^{\circ},\L^{\circ}}(\widetilde{m}\cdot
h_{1},\widetilde{l}\cdot h_{2}) & = &
\psi_{\M^{\circ}}(\widetilde{m})\psi_{\L^{\circ}}^{-1}(\widetilde
{l})\I_{\M^{\circ},\L^{\circ}}(h_{1},h_{2}),
\\
\I_{\M^{\circ},\L^{\circ}}(h_{1}\cdot h,h_{2}\cdot h) & = & \I_{\M^{\circ},\L^{\circ}%
}(h_{1},h_{2}),
\end{eqnarray*}
for every $\widetilde{m} \in\widetilde{\M},
\widetilde{l}\in\widetilde{\L}$ and $h_i\in \E$.

\item \textbf{Normalization}.
$$
\I_{\L^{\circ},\L^{\circ}}(h_{1},h_{2})=\left\{
\begin{array}
[c]{c}%
\psi_{\L^{\circ}}(\widetilde{l})\text{ if }h_{1}=\widetilde{l}\cdot h_{2},\\
0\text{ \ \ \ \ \ \ Otherwise.}%
\end{array}
\right.
$$
for every $\widetilde{l}\in\widetilde{\L}$ and $h_i\in \E$.
\item \textbf{Invariance}.
$$
\I_{g\M^{\circ},g\L^{\circ}}(g(h_{1}),g(h_{2}))=\I_{\M^{\circ},\L^{\circ}}(h_{1},h_{2}),
$$
for every $g\in \G$ and $h_i\in \E$.

\item \textbf{Convolution.}
$$
\I_{\N^{\circ},\M^{\circ}}\ast \I_{\M^{\circ
},\L^{\circ}}=\I_{\N^{\circ},\L^{\circ}},
$$
for every triple $\N^{\circ},\M^{\circ },\L^{\circ}\in
\Lag^{\circ}$. Here $\ast$ is the convolution of kernels:
\[
\I_{\N^{\circ},\M^{\circ}}\ast
\I_{\M^{\circ},\L^{\circ}}(h_{1},h_{2}) :=
{\textstyle\sum\limits_{h\in\widetilde{\M}\backslash \E}}
\I_{\N^{\circ},\M^{\circ}}(h_{1},h)\I_{\M^{\circ},\L^{\circ}}(h,h_{2}).
\]

\item \textbf{Sign Rule.}
$$
\I_{a\M^{\circ},a\L^{\circ}}(h_{1},h_{2}%
)=\chi_{_\mathrm{q}}(a)\I_{\M^{\circ},\L^{\circ}}(h_{1},h_{2}),
$$
for every $a\in \G_{m}$ and $h_i\in \E$.
\end{enumerate}

Note that Property 1 states that the kernel
$\I_{\M^{\circ},\L^{\circ}}$ represents an intertwining operator,
namely, an operator from $\H_{\L^{\circ}} $ to $\H_{\M^{\circ}}$,
which commutes with the action of the Heisenberg group. The other
properties 2-5 are only a reformulation of properties 1-4 from
Theorem \ref{CIO}.

\subsubsection{Diagrammatic description}

Our next step will be to obtain a diagrammatic description of
properties 1-5 above. In order to do this, we have to fix some
additional notations.

In general, when we have a set $\X$, and another set $S$, which is
called a base, we use the notation $\X_{S} :=\X\times S$. We
always (unless explicitly stated otherwise) use the notation
$pr:\X_{S}\rightarrow S$ for the standard projection on the base.

Consider the set $\Lag=\LGr(\V,\omega)$. Let $C\rightarrow
\Lag^{\circ}$ be the pull-back of the canonical vector bundle on
$\Lag$ via the forgetful map $\Lag^{\circ}\rightarrow \Lag$, i.e.,
$C_{\L^{\circ}}=\L$ for every $\L^{\circ}\in \Lag^{\circ}$. We
also define the extended vector bundle $\widetilde {C}\rightarrow
\Lag^{\circ}$ by $\widetilde{C}_{L^{\circ}}=pr^{-1}(L)$, where
$pr:\E\rightarrow \V$ is the standard projection. Finally define
$\psi_{\widetilde{C}}:\widetilde{C}\rightarrow \C$ as
$\psi_{\widetilde{C}}=p^{\ast}\psi$, where $p$ is the total
projection on the center, i.e., $p:\widetilde{C}\rightarrow
Z_{\Lag^{\circ}}=Z\times \Lag^{\circ }\rightarrow Z$.

\textbf{Actions.}

\begin{itemize}
\item Denote by $pr_{1}:\Lag^{\circ}\times \Lag^{\circ}\rightarrow \Lag^{\circ}$
and $pr_{2}:\Lag^{\circ}\times \Lag^{\circ}\rightarrow
\Lag^{\circ}$ the projections on the first and the second
coordinate correspondingly. We define the action:
\begin{equation}
a:(pr_{1}^{\ast}\widetilde{C}\times
pr_{2}^{\ast}\widetilde{C})\times
_{\Lag^{\circ2}}\E_{\Lag^{\circ2}}^{2}\rightarrow
\E_{\Lag^{\circ2}}^{2},\label{action1}
\end{equation}
given (fiberwise) by:
$$
a_{\M^{\circ},\L^{\circ}}(\widetilde{m},\widetilde
{l},h_{1},h_{2})=(\widetilde{m}\cdot h_{1},\widetilde{l}\cdot
h_{2}).
$$

\item We define the action:
\begin{equation}
b:\E_{\Lag^{\circ2}}^{2}\times_{\Lag^{\circ2}}\E_{\Lag^{\circ2}}\rightarrow
\E_{\Lag^{\circ2}}^{2},\label{action2}
\end{equation}
given (fiberwise) by:
$$
b_{\M^{\circ},\L^{\circ}}(h_{1},h_{2},h)=(h_{1}\cdot h,h_{2}\cdot
h).
$$

\item We define the action:
\begin{equation*}
m:\G\times \E_{\Lag^{\circ2}}^{2}\rightarrow
\E_{\Lag^{\circ2}}^{2},
\end{equation*}

given by:
$$
m(g,h_{1},h_{2},\M^{\circ},\L^{\circ})=(gh_{1},gh_{2},g\M^{\circ
},g\L^{\circ}).
$$

\item We define the action:
\begin{equation*}
h:\G_{m}\times \E_{\Lag^{\circ2}}^{2}\rightarrow
\E_{\Lag^{\circ2}}^{2},
\end{equation*}

given by:
$$
h(a,h_{1},h_{2},\M^{\circ},\L^{\circ})=(h_{1},h_{2},a\M^{\circ
},a\L^{\circ}).
$$
\end{itemize}

The collection of kernels $\{\I_{\M^{\circ},\L^{\circ}}\}$ forms a
single function $\I:\E_{\Lag^{\circ2}}^{2}\rightarrow \C$,
satisfying the following properties:

\begin{enumerate}
\item $a^{\ast}\I=\psi_{\widetilde{C}}\boxtimes\psi_{\widetilde{C}}^{-1}\cdot
pr^{\ast}\I$, where $pr=pr_{\E_{\Lag^{\circ2}}^{2}}$.

\item $b^{\ast}\I=pr^{\ast}\I$, where $pr=pr_{\E_{\Lag^{\circ2}}^{2}}$.

\item \textbf{Invariance.}
$$
m^{\ast}\I=pr^{\ast}\I,
$$
where $pr=pr_{\E_{\Lag^{\circ 2}}^{2}}$.

\item \textbf{Convolution.}
$$
pr_{12}^{\ast}\I\ast pr_{23}^{\ast} \I=pr_{13}^{\ast}\I.
$$
In more detail, consider the projections
$$
pr_{ij}:\Lag^{\circ}\times \Lag^{\circ}\times
\Lag^{\circ}\rightarrow \Lag^{\circ}\times \Lag^{\circ},
$$
that are given for $i\neq j$ by
$$
pr_{ij}(\L_{1}^{\circ},\L_{2}^{\circ},\L_{3}^{\circ})=\L_{k}^{\circ},\,k\neq
i,j.
$$
Define:
\begin{equation}
pr_{12}^{\ast}\I\ast
pr_{23}^{\ast}\I=pr_{13_{!}}\widetilde{(\Delta^{\ast
}pr_{12}^{\ast}\I\cdot pr_{23}^{\ast}\I)}\label{convformula}
\end{equation}
where:
$$
\Delta:\E_{\Lag^{\circ2}}^{3}\rightarrow pr_{12}^{\ast}\E_{\Lag^{\circ2}%
}^{2}\times_{\Lag^{\circ2}}pr_{23}^{\ast}\E_{\Lag^{\circ2}},
$$
is the diagonal map:
$$
\Delta_{\M^{\circ},\L^{\circ}}(h_{1},h_{2},h_{3})=(h_{1},h_{2},h_{3}).
$$
The function $\Delta^{\ast}(pr_{12}^{\ast}\I\cdot
pr_{23}^{\ast}\I)$, which lives on the set
$\E_{\Lag^{\circ2}}^{3}$ is $pr_{2}^{\ast}\widetilde{C}$
invariant, so it descends to a function $\ \widetilde{\Delta^{\ast}%
(pr_{12}^{\ast}\I\cdot pr_{23}^{\ast}\I})$ on
$pr_{2}^{\ast}\widetilde {C}\backslash \E_{\Lag^{\circ2}}^{3}$.

We use the notation $pr_{13_{!}}$ for the operation of taking the
sum of values over the fibers of the map $pr_{13}$ (fiberwise
integration). The choice of this notation will become clear when
we translate to the geometric setting.

\item \textbf{Sign Rule. }$h^{\ast}\I=\chi_{_\mathrm{q}}\boxtimes pr^{\ast}\I$, where $pr=$
$pr_{\E_{\Lag^{\circ2}}^{2}}$.
\end{enumerate}

\subsection{Geometric canonical intertwining operators}
In the sequel, we are going to translate back and forth between
algebraic varieties defined over the finite field
$\mathbb{F}_{q}$, and their corresponding sets of rational points.
In order to prevent confusion between the two, we use bold-face
letters for denoting a variety $\mathbf{X}$, and normal letters
for
denoting its corresponding set of rational points $\X=\mathbf{X}(\mathbb{F}%
_{q})$. We recall that an algebraic variety $\mathbf{X}$, namely a
variety in the usual sense - over the algebraically closed field
$\overline{\mathbb F}_q$, is said to be defined over
$\mathbb{F}_{q}$ if it is equipped with the Frobenius endomorphism
$\Fr:\mathbf{X\rightarrow X}$. The set $\X=\mathbf{X}%
(\mathbb{F}_{q})$ is the set of points, fixed by the Frobenius,
$\mathbf{X}(\mathbb{F}_{q})=\mathbf{X}^{\Fr}=\{x\in\mathbf{X}:\,\Fr(x)=x\}$.
The Frobenius structure is also called rational structure. We
denote by $\mathcal{D}^{\w}(\mathbf{X}) =
\mathcal{D}^{\w,b}(\mathbf{X})$ the bounded derived category of
$\ell$-adic Weil sheaves. We choose once an identification
$\Qlb\simeq \C$, so all sheaves are considered over the complex
numbers. Given an object $\SG\in$ $\mathcal{D}^{\w}(\mathbf{X})$,
one can associate to $\SG$ a function
$f^{\SG}:\X\rightarrow \C$, as follows:%
\[
f^{\SG}(x):=%
{\textstyle\sum\limits_{i}}
(-1)^{i}\,\Tr(\Fr_{|\mathrm{H}^{i}(\SG_{x})}).
\]
This procedure is called Grothendieck's {\it sheaf-to-function
correspondence} \cite{Gr}. We also use the notation
$\chi_{_\Fr}(\mathcal{G}) := f^{\SG}$, and call it the
\textit{Euler characteristic} of the sheaf $\SG$. Hence, we can
start geometrizing our constructions.

The symplectic space $(\V,\omega)$ can be naturally identified as
the set $\V=\mathbf{V}(\mathbb{F}_{q})$, where $\mathbf{V}$ is an
algebraic variety, defined over $\mathbb{F}_{q}$,
$\mathbf{V}\simeq\mathbb{A}^{2\n}$, equipped
with a skew symmetric form $\mathbf{\omega}:\mathbf{V \times V\rightarrow}%
\mathbb{A}^{1}$, respecting the rational structure of both sides.
We have $\omega=\mathbf{\omega}_{|V}$. The Heisenberg group $\E$,
can be naturally
identified as the set $\E=\mathbf{H}(\mathbb{F}_{q})$, where $\mathbf{H}%
=\mathbf{V\times}\mathbb{A}^{1}$ with the same multiplication
formulas. We have the center
$\mathbf{Z}=Z(\mathbf{H})=\{(0,z):z\in \mathbb{A}^{1}\}$. We have
the Artin-Schreier sheaf $\mathscr{L}_{\psi}$ associated with the
central character $\psi:Z\rightarrow \C^{\ast}$, that is,
$f^{\mathscr{L}_{\psi}}=\psi$. The group $\G=\Sp(\V,\omega)$
is identified as $\G=\mathbf{G}(\mathbb{F}_{q})$, where $\mathbf{G}%
=\Sp(\mathbf{V,\omega})$. Next, we replace the sets $\Lag^{\circ},
C$ and $\widetilde {C}$ by the corresponding algebraic varieties
$\mathbf{Lag}^{\circ
}, \mathbf{C} $ and $\widetilde{\mathbf{C}}$. We have the sheaf $\mathscr{L}%
_{\psi,\widetilde{C}}=p^{\ast}\mathscr{L}_{\psi}$, where
$p:\widetilde {C}\rightarrow\mathbf{Z}$ is the total projection on
the center. The sheaf $\mathscr{L}_{\psi,\widetilde{C}}$ is
associated via sheaf-to-function
correspondence to the function $\psi_{\widetilde{C}}$ , $f^{\mathscr{L}%
_{\psi,\widetilde{C}}}=\psi_{\widetilde{C}}$. Finally, we can
define the actions $a,b,m$ and $h$ between the corresponding
algebraic varieties. All actions respect rational structure, and
reduce to give the old formulas between sets of rational points.
We have the following fundamental theorem:

\begin{theorem}[Geometric canonical intertwining operators]\label{gcio}
There exists a unique $($up to unique isomorphism $)$
$($shifted$)$ Weil perverse
sheaf $\SF$ on the variety $\mathbf{H}_{\mathbf{Lag}%
^{\circ2}}^{2}$, equipped with the following data:

\begin{enumerate}
\item $a^{\ast}\SF\simeq\mathscr{L}_{\psi,\widetilde{C}}%
\mathscr{\boxtimes L}_{\psi^{-1},\widetilde{C}}\otimes
pr^{\ast}\SF$, where
$pr=pr_{\mathbf{H}_{\mathbf{Lag}^{\circ2}}^{2}}.$

$b^{\ast}\SF\simeq pr^{\ast}\SF$, where $pr=pr_{\mathbf{H}%
_{\mathbf{Lag}^{\circ2}}^{2}}$.

\item \textbf{Normalization.} $\Delta^{\ast}\SF\simeq (\C)_{\mathbf{Lag}^{\circ}},$
where $\Delta:\mathbf{Lag}^{\circ
}\rightarrow\mathbf{H}_{\mathbf{Lag^{\circ}}}^{2}, \,
\Delta(\mathbf{L}^{\circ
})=(\mathbf{L}^{\circ},\mathbf{L}^{\circ},0,0)$, and $0$ denote
the unit element in $\mathbf{H}$.

\item \textbf{Equivariance. }$m^{\ast}\mathcal{F\simeq}pr^{\ast}\SF$,
where $pr=pr_{\mathbf{H}_{\mathbf{Lag^{\circ}}}^{2}}$.

\item \textbf{Convolution. }$pr_{12}^{\ast}\SF\ast pr_{23}^{\ast
}\SF\simeq pr_{13}^{\ast}\SF$, where%
\[
pr_{12}^{\ast}\SF\ast pr_{23}^{\ast}\mathcal{F=}\bigskip\ pr_{13_{!}%
}(\widetilde{\Delta^{\ast}(pr_{12}^{\ast}\SF\otimes pr_{23}^{\ast
}\SF})).
\]
We use the same terminology as in $\mathrm{(\ref{convformula})}$,
noting that the sheaf $\Delta^{\ast}(pr_{12}^{\ast}\SF\otimes
pr_{23}^{\ast}\mathcal{F})$ is $pr_{2}^{\ast}\widetilde{C}$
equivariant so it descent to a sheaf
$\widetilde{\Delta^{\ast}(pr_{12}^{\ast}\SF\otimes pr_{23}^{\ast
}\SF})$ on $pr_{2}^{\ast}\widetilde{C}\backslash\mathbf{H}%
_{\mathbf{Lag^{\circ}}}^{3}$.

\item \textbf{Sign rule. }$h^{\ast}\SF\simeq\mathscr{L}_{\chi_{_\mathrm{q}}
}\boxtimes pr^{\ast}\SF$, where
$pr=pr_{\mathbf{H}_{\mathbf{Lag^{\circ }}}^{2}}$, and
$\mathscr{L}_{\chi_{_\mathrm{q}}}$ denotes the Kummer sheaf on
$\mathbb{G}_{m}$ associated to the quadratic character
$\chi_{_\mathrm{q}}$.
\end{enumerate}
The isomorphisms 1-5 satisfy some obvious compatibility conditions
which we will omit here.
\end{theorem}
\subsection{Geometric models}
We fix an element $\mathbf{L^{\circ}\in Lag^{\circ}}$ of the form $\mathbf{L^{\circ}%
=(L,}\sigma_{\mathbf{L}})$. We denote by $\mathcal{D}(\mathbf{H)}
:= \mathcal{D}^b(\mathbf{H)}$ the bounded derived category of
$\ell$-adic sheaves on $\mathbf{H}$.

Recall the notation $\widetilde{\mathbf{L}}=pr^{-1}(\mathbf{L})$,
where
$pr:\mathbf{H\rightarrow V}$ is the standard projection. Consider the sheaf $\mathscr{L}%
_{\psi,\mathbf{{L^{\circ}}}} :=p^{\ast}\mathscr{L}_{\psi}$, where
$p:\widetilde {\mathbf{L}}=\mathbf{L\times
Z}\rightarrow\mathbf{Z}$.

We denote by ${\mathcal
D}_{\mathbf{L^{\circ}}}=\mathcal{D}^b_{\mathbf{L^{\circ},\psi}}(\mathbf{H})$
the bounded derived category of
$\mathscr{L}_{\psi,\mathbf{L^{\circ}}} $-equivariant sheaves. In
more detail, the subgroup $\widetilde{\mathbf{L}}
\subset\mathbf{H}$ acts on the group $\mathbf{H}$ via left
multiplication. We
denote this action by $a_{\mathbf{L^{\circ}}}:\widetilde{\mathbf{L}}%
\times\mathbf{H\rightarrow H}$ and we define:
\\
\begin{definition}
An $\mathscr{L}_{\psi,\mathbf{L^{\circ}}} $-equivariant structure
on a sheaf $\SG \in \mathcal{D}_{\mathbf{L^{\circ}}}$ is given by
an isomorphism:
$\alpha:a_{\mathbf{L^{\circ}}}^{\ast}\mathcal{G\simeq}\mathscr{L}%
_{\psi,\mathbf{L^{\circ}}}\boxtimes\SG$, satisfying a cocycle
condition.
\end{definition}

\textbf{Comment.} Note that the subgroup $\widetilde{\mathbf{L}}$
acts freely on $\mathbf{H}$. Hence, in the definition of
$\mathcal{D}_{\mathbf{L^{\circ}}}$ we do not have to use the
elaborate notion of a "real" equivariant derived category
\cite{BL}.
\\
\\
The triangulated category $\mathcal{D}_{\mathbf{L^{\circ}}}$
admits a perverse
t-structure, and we denote its heart by $\Perv_{\mathbf{L^{\circ}}%
}:=\Perv_{\mathbf{L^{\circ},}\psi}(\mathbf{H})$ - the abelian category of $\mathscr{L}%
_{\psi,\mathbf{L^{\circ}}}$-equivariant perverse sheaves.

\subsubsection{Intertwining functors}

Given a pair $(\mathbf{M^{\circ},L^{\circ})\in Lag^{\circ}\times Lag^{\circ}}%
$, we consider the sheaf $\SF_{\mathbf{M^{\circ},L^{\circ}}}$ -
the
fiber of the sheaf $\SF$ at the point $(\mathbf{M^{\circ},L^{\circ})}%
$. Using the sheaf $\SF_{\mathbf{M^{\circ},L^{\circ}}}$ we
construct a functor that for simplicity we will denote by the same
notation,:
$$
\SF_{\mathbf{M^{\circ},L^{\circ}}}:\mathcal{D}_{\mathbf{L}^{\circ}}\rightarrow
\mathcal{D}_{\mathbf{M}^{\circ}},
$$
given by convolution:
$$
\mathcal{G}\mapsto \SF_{\mathbf{M^{\circ},L^{\circ}}}\ast\SG,
$$
where $\mathcal{G\in D}_{\mathbf{L^{\circ}}}$. It is an important
fact, which follows from a deep theorem of {\it Katz-Laumon} on
the $\ell-$adic Fourier transform \cite{KL}, that the functor
$\SF_{\mathbf{M^{\circ},L^{\circ }}}$ respects the perverse
t-structure on both sides, $\SF_{\mathbf{M^{\circ
},L^{\circ}}}(\Perv_{\mathbf{L^{\circ}}})\subset
\Perv_{\mathbf{M}^{\circ}}$.

\subsubsection{Group actions}
We have actions of the groups $\mathbf{H}$ and $\mathbf{G}$ on the
category $\mathcal{D}_{\mathbf{L^{\circ}}}$. These constitute a
categorification of the Heisenberg representation $\pi$, and the
Weil representation $\rho$ correspondingly.

\textbf{Heisenberg action. } The Heisenberg group $\mathbf{H}$
acts on itself by right multiplication. We denote this action by
$R:\mathbf{H\times H\rightarrow H}$. This induces an action of
$\mathbf{H}$ on the category $\mathcal{D}_{\mathbf{L^{\circ}}}$.
Namely, associated to every element $h\in \mathbf{H}$ we have a
functor $\SK_{\mathbf{\pi}}(h) = \SK_{\pi \mathbf{,L^{\circ}}}(h)
: \mathcal{D}_{\mathbf{L^{\circ}}}\longrightarrow
\mathcal{D}_{\mathbf{L^{\circ}}}$ given by:
\begin{eqnarray*}
\SK_{\mathbf{\pi}}(h) &:=&  R_{h}^{\ast},
\end{eqnarray*}
the pull-back by the map $R_{h}:\mathbf{H\rightarrow H}$.

\begin{claim}
We have a canonical isomorphism of functors:
$$
\SK_{\mathbf{\pi}}(h_{1})\circ\SK
_{\mathbf{\pi}}(h_{2})\simeq\SK_{\mathbf{\pi}}(h_{1}\cdot h_{2}),
$$
for every $h_{1},h_{2}\in\mathbf{H}$.

\end{claim}
\textbf{Symplectic action.} The symplectic group $\mathbf{G}$ acts
on $\mathbf{H}$ via its tautological action on the vector space
$\mathbf{V}$. This induces an action of $\mathbf{G}$ on the
category $\mathcal{D}(\mathbf{H})$. In more detail, associated to
every element $g\in\mathbf{G}$ we have a functor $\widetilde
{\SK}_{\mathbf{\rho}}(g):\mathcal{D}(\mathbf{H})\rightarrow
\mathcal{D}(\mathbf{H})$, given by:
$$
\widetilde{\SK}_{\mathbf{\rho}}(g)(\SG)=(g^{-1})^{\ast}%
\SG.$$ We have a canonical isomorphism of functors:
\begin{equation}
\widetilde{\SK}_{\mathbf{\rho}}(g_{1})\circ\widetilde{\SK}%
_{\mathbf{\rho}}(g_{2})\simeq\widetilde{\SK}_{\mathbf{\rho}}(g_{1}\cdot
g_{2})\label{iso1}%
\end{equation}
for every $g_{1},g_{2}\in\mathbf{G}$.

However, the functor $\widetilde{\SK
}_{\mathbf{\rho}}(g)$ does not preserve the category $\mathcal{D}%
_{\mathbf{L^{\circ}}}$. In fact, we have $\widetilde{\SK}_{\mathbf{\rho}%
}(g):\mathcal{D}_{\mathbf{L}^{\circ}}\rightarrow \mathcal{D}_{g\mathbf{L}%
^{\circ}}$. In order to arrive back to the category $\mathcal{D}%
_{\mathbf{L^{\circ}}}$, we apply next the intertwining functor
$\SF _{\mathbf{L^{\circ},}g\mathbf{L^{\circ}}}$, so we define $\SK
_{\mathbf{\rho}}(g)=\SK_{\rho\mathbf{,L^{\circ}}}(g):\mathcal{D}%
_{\mathbf{L^{\circ}}}\rightarrow
\mathcal{D}_{\mathbf{L^{\circ}}}$, by:
$$
\SK_{\mathbf{\rho}}(g):=\SF_{\mathbf{L^{\circ},}g\mathbf{L^{\circ}}%
}\circ\widetilde{\SK}_{\mathbf{\rho}}(g).
$$
We have an isomorphism:
$$
\SK_{\rho}(g_{1})\circ\SK_{\mathbf{\rho}}(g_{2})\simeq\SK
_{\mathbf{\rho}}(g_{1}\cdot g_{2}),
$$
for every $g_{1},\; g_{2}\in\mathbf{G}$. The last isomorphism is a
formal consequence of isomorphism (\ref{iso1}), and the
convolution property of the sheaf of intertwining kernels $\SF$
(see Property 4, in Theorem \ref{gcio}).

\subsection{Deligne's Weil representation sheaf}\label{Deligne's sheaf}
The correct point of view is to construct a single sheaf,
$\mathcal{K}_{\mathbf{\rho}}=\mathcal{K}_{\rho,\mathbf{L^{\circ}}}$,
on the variety $\mathbf{G\times H\times H}$. Consider the map:
$$
m_{\mathbf{L^{\circ
}}}:\mathbf{G\times H\times H\rightarrow\mathbf{H}_{\mathbf{Lag^{\circ}}}^{2}%
},
$$
given by:
$$
m_{\mathbf{L^{\circ}}}(g,h_{1},h_{2})=(h_{1},g(h_{2}
),\mathbf{L^{\circ},}g\mathbf{L^{\circ}}).
$$
We define:
$$\mathcal{K}_{\mathbf{\rho}} :=
m_{\mathbf{L^{\circ}}}^{\ast}\SF,
$$
where $\SF$ is the sheaf of intertwining kernels (cf. Theorem
\ref{gcio}).

Now we have the following actions:
$$
a:(\widetilde{\mathbf{L}}_{\mathbf{G}}\times_{\mathbf{G}%
}\widetilde{\mathbf{L}}_{\mathbf{G}})\times_{\mathbf{G}}\mathbf{H}%
_{\mathbf{G}}^{2}\rightarrow\mathbf{H}_{\mathbf{G}}^{2},
$$
given (fiberwise) by:
$$
a_{g}(\widetilde{l}_{1},\widetilde{l}_{2},h_{1},h_{2}) := (\widetilde{l}%
_{1}\cdot h_{1},\widetilde{l}_{2}\cdot h_{2}),
$$
and:
$$b:\mathbf{\mathbf{H}%
_{\mathbf{G}}^{2}\times}_{\mathbf{G}}\mathbf{\mathbf{H}_{\mathbf{G}%
}\rightarrow\mathbf{H}_{\mathbf{G}}^{2}},
$$
given (fiberwise) by:
$$
b_{g} (h_{1},h_{2},h):=(h_{1}\cdot h,h_{2}\cdot h).
$$
Note that we use the same notations as in (\ref{action1}), and
(\ref{action2}), because we deal here essentially with the same
actions.

\begin{claim}[Weil representation sheaf] The sheaf $\mathcal{K}_{\mathbf{\rho}}$ satisfies the
following properties:
\end{claim}
\begin{enumerate}
\item $a^{\ast}\mathcal{K}_{\mathbf{\rho}}\simeq(\mathscr{L}_{\psi
,\mathbf{L_{\mathbf{G}}^{\circ}}}\boxtimes\mathscr{L}_{\psi^{-1}%
,\mathbf{L_{G}^{\circ}}})\boxtimes\mathcal{K}_{\mathbf{\rho}}$, $\
$where
$\mathscr{L}_{\psi,\mathbf{L_{G}^{\circ}}}=p^{\ast}\mathscr{L}_{\psi}$,
$p:\widetilde{\mathbf{L}}_{\mathbf{G}}=\widetilde{\mathbf{L}}\times
\mathbf{G\rightarrow}\widetilde{\mathbf{L}}\rightarrow\mathbf{Z}$.

$b^{\ast}\mathcal{K}_{\mathbf{\rho}}\simeq
pr^{\ast}\SF_{\mathbf{\rho }}$, where
$pr=pr_{\mathbf{\mathbf{H}_{\mathbf{G}}^{2}}}$.

\item \textbf{Convolution. }$pr_{1}^{\ast}\mathcal{K}_{\mathbf{\rho}}\ast
pr_{2}^{\ast}\mathcal{K}_{\mathbf{\rho}}\simeq m^{\ast}\mathcal{K}%
_{\mathbf{\rho}}$. In more detail, $pr_{1},pr_{2}:\mathbf{G\times
G\rightarrow G}$ are projections on the left and right coordinate
correspondingly, $m:\mathbf{G\times G\rightarrow G}$ is the
multiplication map. Considering the following sequence of maps:
\[
pr_{1}^{\ast}\mathbf{H}_{\mathbf{G}}^{2}\times_{\mathbf{G\times G}}%
pr_{2}^{\ast}\mathbf{H}_{\mathbf{G}}^{2}\overset{\Delta}{\leftarrow}%
\mathbf{H}_{\mathbf{G}}^{3}\twoheadrightarrow\widetilde{\mathbf{L}%
}_{\mathbf{G}}\backslash\mathbf{H}_{\mathbf{G}}^{3}\overset{pr_{13}%
}{\rightarrow}m^{\ast}\mathbf{H}_{\mathbf{G}}^{2}%
\]
We define $pr_{1}^{\ast}\mathcal{K}_{\mathbf{\rho}}\ast
pr_{2}^{\ast
}\mathcal{K}_{\mathbf{\rho}}=pr_{13_{!}}(\widetilde{\Delta^{\ast}(pr_{1}%
^{\ast}\mathcal{K}_{\mathbf{\rho}}\ast
pr_{2}^{\ast}\SF_{\mathbf{\rho }})})$.
\end{enumerate}

\textbf{Comment.} The sheaf $\mathcal{K}_{\mathbf{\rho}}$ should
be considered as a sheaf of kernels, which constitutes a geometric
analogue of the Weil representation $\rho$, realized in \ a model
associated with the enhanced Lagrangian $\mathbf{L^{\circ}}$.
Indeed, in the case where $\mathbf{L^{\circ}}$ is
rational, namely $\Fr \,\mathbf{L^{\circ}=L^{\circ}}$, the sheaf $\mathcal{K}%
_{\mathbf{\rho}}$ admits a Frobenius (Weil) structure, and taking
the corresponding function $f^{\mathcal{K}_{\mathbf{\rho}}}$ gives
the kernels of the representation
$(\rho_{\L^{\circ}},\G,\H_{\L^{\circ}})$, $\L^{\circ
}=\mathbf{L^{\circ}(}\mathbb{F}_{q})$. We would like to complement
that $\mathcal{K}_{\mathbf{\rho}}$ is the sheaf appearing in
Deligne's letter to Kazhdan from 1982 \cite{D1, GH3}.

\subsection{Twisted Weil structure}
Consider the Frobenius endomorphism
$\Fr=\Fr_{\mathbf{H}}:\mathbf{H\rightarrow H}$. Assume we are
given an object ${\SG} \in \mathcal{D}_{\mathbf{L}^{\circ},\psi}$.
Then we have:
$$
\Fr^{\ast}\SG \in \mathcal{D}_{\Fr^{-1}\mathbf{L}^{\circ},\psi},
$$
where $\Fr^{-1}\mathbf{L^{\circ}%
=(}\Fr^{-1}\mathbf{L,}\Fr^{-1}\sigma_{\mathbf{L}})$. We use here
the fact that
$\mathscr{L}_{\psi}$ is a Weil sheaf, $\Fr^{\ast}\mathscr{L}_{\psi}%
\simeq\mathscr{L}_{\psi}$.

Now, in order to arrive back to the category
$\mathcal{D}_{\mathbf{L}^{\circ},\psi}$, we apply next the
intertwining functor to obtain:
$$
\SF_{\mathbf{L^{\circ},}\Fr^{-1}\mathbf{L}^{\circ}}(\Fr^{\ast
}\SG)\in \mathcal{D}_{\mathbf{L}^{\circ},\psi}.
$$
In fact, this procedure defines a functor
$$
\Fr_{\mathbf{L^{\circ}}}:=\SF_{\mathbf{L^{\circ},}\Fr^{-1}\mathbf{L}
^{\circ}}\circ
\Fr^{\ast}:\mathcal{D}_{\mathbf{L}^{\circ},\psi}\rightarrow
\mathcal{D}_{\mathbf{L}^{\circ},\psi}.
$$
Moreover, we have
$\Fr_{\mathbf{L^{\circ}}}(\Perv_{\mathbf{L^{\circ},\psi}})\subset
\Perv_{\mathbf{L^{\circ},\psi}}$. The functor
$\Fr_{\mathbf{L^{\circ}}}$ is thought of as a generalized
Frobenius.

We define the triangulated category
$\mathcal{D}_{\mathbf{L}^{\circ},\psi}^{\w}=\mathcal{D}^{\w}\mathcal{(}%
\mathbf{H,L^{\circ},}\psi)$ of $\mathbf{L^{\circ}}$- Weil sheaves.
An object
in $\mathcal{D}_{\mathbf{L}^{\circ},\psi}^{\w}$ is a pair $\mathcal{(F}%
,\alpha)$, where $\mathcal{G\in D}_{\mathbf{L}^{\circ},\psi}$, and
$\ \alpha$
is an isomorphism $\alpha:\Fr_{\mathbf{L^{\circ}}}^{\ast}\SG%
\simeq\SG$. The category
$\mathcal{D}_{\mathbf{L}^{\circ},\psi}^{\w}$
inherits a perverse t-structure. We denote by $\Perv_{\mathbf{L^{\circ},\psi}%
}^{\w}=\Perv^{\w}(\mathbf{H,L^{\circ},}\psi)$ the abelian heart of
perverse $\mathbf{L^{\circ}}$-Weil sheaves. We often use the
simplified notation $\mathcal{D}_{\mathbf{L}^{\circ}}^{\w}$,
$\Perv_{\mathbf{L^{\circ}}}^{\w}$, assuming $\psi$ is known from
the context.

Given a pair $(\mathbf{M^{\circ},L^{\circ})}$, the intertwining
functor $\SF_{\mathbf{M}^{\circ}\mathbf{,L}^{\circ}}$ respects the
Weil structures on both sides, therefore it gives a functor
between corresponding Weil categories:
$\mathcal{D}_{\mathbf{L}^{\circ}}^{\w}$ and $\mathcal{D}_{\mathbf{M^{\circ}}}^{\w}%
$. We denote this functor also by $\SF_{\mathbf{M}^{\circ}\mathbf{,L}%
^{\circ}}: \mathcal{D}_{\mathbf{L}^{\circ}}^{\w}\rightarrow \mathcal{D}%
_{\mathbf{M^{\circ}}}^{\w}$. We have $\SF_{\mathbf{M}^{\circ}\mathbf{,L}%
^{\circ}}(\Perv_{\mathbf{L^{\circ}}}^{\w})\subset
\Perv_{\mathbf{M^{\circ}}}^{\w}$.

\textbf{Comment.} In the case where $\mathbf{L^{\circ}}$ is
rational, i.e., $\Fr\,\mathbf{L^{\circ}=L^{\circ}}$ we have
$\SF_{\mathbf{L^{\circ},}\Fr^{-1}\mathbf{L}^{\circ}}=\mathrm{Id}$.
Hence, the category $\mathcal{D}_{\mathbf{L}^{\circ}}^{\w}$
becomes only the category of traditional Weil sheaves.

The action functors $\SK_{\pi}$ and $\SK_{\rho}$ restrict to give
actions of the finite groups $\E$ and $\G$ on
$\mathcal{D}_{\mathbf{L}^{\circ}}^{\w}$. The intertwining functor
$\SF_{\mathbf{M}^{\circ}\mathbf{,L}^{\circ}}$ commutes with these
actions.

\subsection{From a category to an Hilbert space}

We denote by $K_{\mathbf{L^{\circ}}}$ and
$K_{\mathbf{L^{\circ}}}^{\w}$ the Grothendieck $K$ groups of the
categories $\mathcal{D}_{\mathbf{L}^{\circ}}$ and
$\mathcal{D}_{\mathbf{L}^{\circ}}^{\w}$ correspondingly. These are
infinite dimensional (complex) vector spaces (at least after
tensoring with $\C$). In particular, the actions $\SK_{\pi}$ and
$\SF_{\rho}$ of the
groups $\E$ and $\G$ factor through, and give the vector space $K_{\mathbf{L^{\circ}}%
}^{\w}$ a structure of an infinite dimensional representation. The
same argument works for the intertwining functor
$\SF_{\mathbf{M}^{\circ }\mathbf{,L}^{\circ}}$, giving an
intertwining homomorphism $\SF
_{\mathbf{M}^{\circ}\mathbf{,L}^{\circ}}:K_{\mathbf{L^{\circ}}}^{\w}\rightarrow
K_{M\mathbf{^{\circ}}}^{\w}$, for every pair
$(\mathbf{L^{\circ},M^{\circ}})$.

Next we are going to define a canonical finite dimensional
quotient. Considering $\mathcal{G\in}$
$\mathcal{D}_{\mathbf{L}^{\circ},\psi}^{\w}$, we have
$\mathbb{D}(\mathcal{G)\in D}_{\mathbf{L}^{\circ},\psi^{-1}}$,
where we denote by $\mathbb{D}$ the functor of Vardier duality.

\textbf{Matrix coefficient. }Let $pr_{1},pr_{2}:\mathbf{H\times H
\rightarrow H}$, be the projections on the left and right
coordinates correspondingly. Let $R:\mathbf{H\times H\rightarrow
H}$ denote the right action of $\mathbf{H}$ on itself. Given a
pair $\SE,\mathcal{G\in D}_{\mathbf{L}^{\circ}}^{\w}$. Considering
the product $pr_{1}^{\ast}\mathbb{D}(\SE)\otimes r^{\ast
}\mathcal{G}$, it is an equivariant sheaf  with respect to the
diagonal action of $\widetilde{\mathbf{L}}$ on $\mathbf{H\times
H}$, therefore it descends to a sheaf
$\widetilde{pr_{1}^{\ast}\mathbb{D}(\SE)\otimes
r^{\ast}\mathcal{G}}$ on the quotient
$\widetilde{\mathbf{L}}\backslash(\mathbf{H\times H)}$. Define
the sheaf:%
\begin{equation}
\m(\SE,\mathcal{G)}:=pr_{2_{!}}(\widetilde{pr_{1}^{\ast}\mathbb{D}%
(\SE)\otimes r^{\ast}\mathcal{G}})\label{matrix}%
\end{equation}
The sheaf $\m(\SE,\mathcal{G)}$ is called the \textit{matrix
coefficient} of $\SE$ and $\mathcal{G}$. We have the following
fundamental statement:

\begin{proposition}
The sheaf $\m(\SE,\mathcal{G)}$ admits a Frobenius structure,
namely $\m(\SE,\mathcal{G)}$ $\in
\mathcal{D}^{\mathrm{\w}}(\mathbf{H})$.
\end{proposition}

Using the previous result, we next apply sheaf-to-function
correspondence
obtaining a function $f^{\m(\SE,\mathcal{G)}}:\E\rightarrow \C$. To a pair $\SE$, $\mathcal{G}\in$ $\mathcal{D}%
_{\mathbf{L}^{\circ}}^{\w}$ we can associate a function $f^{\m(\SE%
,\mathcal{G)}}$ on $\E$.

This procedure clearly factorizes to the level of $K$ groups,
giving a $(\E,\G)$ invariant pairing:
\begin{equation*}
\left\langle \cdot,\cdot\right\rangle _{\mathbf{L^{\circ}}}%
:K_{\mathbf{L^{\circ}}}^{\w}\times
K_{\mathbf{L^{\circ}}}^{\w}\rightarrow \L^{2}(\E,\C).
\end{equation*}

Denoting by $N_{\mathbf{L^{\circ}}}^{\w}\subset
K_{\mathbf{L^{\circ}}}^{\w}$ the radical of $\left\langle
\cdot,\cdot\right\rangle _{\mathbf{L^{\circ}}}$, we define
$\H_{\mathbf{L^{\circ}}}=K_{\mathbf{L^{\circ}}}^{\w}/N_{\mathbf{L^{\circ
}}}^{\w}$. Following the work \cite{BP} we will call the quotient
map:
\begin{equation}\label{BPC}
K_{\mathbf{L^{\circ}}}^{\w} \twoheadrightarrow
\H_{\mathbf{L^{\circ}}},
\end{equation}
which assigns to a sheaf $\SG$ the \textit{vector} $[\SG] \in
\H_{\mathbf{L^{\circ}}}$, the \textit{Braverman-Polishchuk's
Correspondence}\footnote{This correspondence becomes the usual
Grothendieck's sheaf-to-function correspondence in the case where
the enhanced Lagrangian $\mathbf{L^{\circ}}$ is rational.}.

This correspondence is compatible with the actions of the groups
$\E$ and $\G$, hence the vector space $\H_{\mathbf{L^{\circ}}}$
inherits the actions
of $\E$ and $\G$ also. We denote these representations by $(\pi_{_\mathbf{L^{\circ}}%
},\E,\H_{\mathbf{L^{\circ}}}),$ and $(\rho_{_\mathbf{L^{\circ}}}%
,\G,\H_{\mathbf{L^{\circ}}})$. We have the following theorem:

\begin{theorem}
The vector space $\H_{\mathbf{L^{\circ}}}$ is finite dimensional.
Moreover,

\begin{enumerate}
\item The representation $(\pi_{_\mathbf{L^{\circ}}},\E,\H_{\mathbf{L^{\circ}}})$
is isomorphic to the Heisenberg representation.

\item The representation $(\rho_{_\mathbf{L^{\circ}}},\G,\H_{\mathbf{L^{\circ}}%
})$ is isomorphic to the Weil representation.
\end{enumerate}
\end{theorem}

Finally, we have that
$\SF_{\mathbf{M}^{\circ}\mathbf{,L}^{\circ}}$ sends
$N_{\mathbf{L^{\circ}}}^{\w}$ to $N_{\mathbf{M^{\circ}}}^{\w}$,
giving an intertwining operator between the models
$\H_{\mathbf{L^{\circ}}}$ and $\H_{\mathbf{M^{\circ}}}$.
\\
\\
\textbf{Summary.} To every enhanced Lagrangian
$\mathbf{L^{\circ}}$, not necessarily rational, we associate a
model $\H_{\mathbf{L^{\circ}}}$ of the Heisenberg-Weil
representations. Underlying every such model lies a category
$\mathcal{D}_{\mathbf{L}^{\circ}}^{\w}$ of (twisted) Weil sheaves.

\textbf{Comment.}
In the case where $\mathbf{L^{\circ}}$ is rational, i.e., $\Fr \,\mathbf{L^{\circ}=L^{\circ}}%
$, we get $\H_{\mathbf{L^{\circ}}}=\H_{\L\mathbf{^{\circ}}}$,
where $\L^{\circ }=\mathbf{L^{\circ}(}\mathbb{F}_{q})$, and
$\H_{\L\mathbf{^{\circ}}}$ is the vector space constructed in
Section \ref{models}.
\section{Hecke Eigenvectors}
In the remainder of this section we consider the case $\n=2$. Let
$(\V,\ome)$ be a two-dimensional symplectic vector space over the
finite field $\Fq$. We consider a maximal torus $\T \subset \G :=
\Sp(\V,\ome)$, which we will call the Hecke torus, and a
multiplicative character $\chi:\T \rightarrow \C^{\ast}$. Our goal
is to construct effectively a $\chi$-Hecke-eigenvector
$v_{\chi}\in \H_{\V}$. We denote by $\mathscr{L}_{\chi}$ the
\textit{Kummer} sheaf on the variety $\mathbf{T}$ associated to
the character $\chi$.

\subsection{Construction}
The idea behind the construction is to use a "good realization"
$\H_{\mathbf{L^{\circ }}}$ in which the Hecke torus acts in a
geometric fashion and to construct first an Hecke eigensheaf.
Then, the second step is to apply the Braverman-Polishchuk
correspondence (\ref{BPC}) to obtain the desired eigenvector.

\subsubsection{Good realization}
There exists an enhanced Lagrangian $\mathbf{L^{\circ}=(L,}\sigma_{\mathbf{L}%
})$ such that $\mathbf{L}$ is fixed by $\mathbf{\T}$, that is,
$g\mathbf{L=L}$ for every $g\in\mathbf{\T}$. Applying the
Frobenius we obtain another enhanced Lagrangian
$\mathbf{L}^{\circ\prime
}=(\mathbf{L}^{\prime},\sigma_{\mathbf{L}^{\prime}}) :=(\Fr \,\mathbf{L,}%
\Fr \, \sigma_{\mathbf{L}})$. We can choose
$\mathbf{L^{\circ\prime}}$ such that
$\omega(\sigma_{\mathbf{L}},\sigma_{\mathbf{L}^{\prime}})=1$. We
consider the
decompositions: $\mathbf{V=L}^{\prime}\mathbf{\oplus L}$ and $\mathbf{H=L}%
^{\prime}\times\mathbf{L\times Z}$. Note, that
$\mathbf{Z=Z(H)=}\mathbb{A}^{1}$.

We have $\mathbf{X}_{\mathbf{L}^{\circ}} :
=\widetilde{\mathbf{L}}\backslash
\mathbf{H}\simeq\mathbf{L}^{\prime}$, therefore we can identify:
\begin{equation}\label{iden1}
\mathcal{D}_{\mathbf{L}^{\circ},\psi} \simeq
\mathcal{D}(\mathbf{L}^{\prime}), \r
\Perv_{\mathbf{L^{\circ},\psi}} \simeq \Perv(\mathbf{L}^{\prime
}).
\end{equation}
Next, we compute the twisted Frobenius $\Fr_{\mathbf{L^{\circ}}}$
in the realization $\mathcal{D}(\mathbf{L}^{\prime})$:
\begin{align*}
\Fr_{\mathbf{L^{\circ}}}(\mathcal{G)(}l^{\prime})  & =\Fr_{\mathbf{L^{\circ}}%
}(\mathcal{G)(}l^{\prime},0,0)\\
& =\SF_{\mathbf{L^{\circ},}\Fr^{-1}\mathbf{L}^{\circ}}\circ
\Fr^{\ast
}\mathcal{G(}l^{\prime},0,0)\\
& =\star%
{\textstyle\int\limits_{l\in\mathbf{L}}}
\Fr^{\ast}\SG((0,l,0)\cdot(l^{\prime},0,0))\\
& = \star%
{\textstyle\int\limits_{l\in\mathbf{L}}}
\Fr^{\ast}\SG((l^{\prime},l,%
\half
\omega(l.l^{\prime}))\\
& =\star%
{\textstyle\int\limits_{l\in\mathbf{L}}}
\SG((\Fr\,l^{\prime},\Fr\,l,%
\half
\Fr\,\omega(l.l^{\prime}))\\
& \above{(1)}{=}\star%
{\textstyle\int\limits_{l\in\mathbf{L}}}
\mathscr{L}_{\psi}(%
\half
\omega(l,l^{\prime})+%
\half
\Fr\omega(l,l^{\prime}))\otimes\mathcal{G(}\Fr\,l)\\
& \above{(2)}{=}%
{\textstyle\int\limits_{l\in\mathbf{L}}}
\mathscr{L}_{\psi}(\omega(l,l^{\prime}))\otimes\mathcal{G(}\Fr\,l)
\end{align*}

In (1) we use the identity: $(\Fr\,l,\Fr\,l^{\prime},%
\half
\omega(l,l^{\prime}))=(0,\Fr\,l^{\prime},%
\half
\omega(l,l^{\prime})-%
\half
\Fr\,\omega(l,l^{\prime}))\cdot(\Fr\,l,0,0)$, and in (2) we use
the fact that
$\mathscr{L}_{\psi}$ is a Frobenius sheaf, i.e., $\Fr^{\ast}\mathscr{L}_{\psi}%
\simeq\mathscr{L}_{\psi}$. The symbol $\star$ stands for a
normalization coefficient which appears in the canonical
intertwining operator. This coefficient, although it has a deep
meaning, does not play any role in our arguments, therefore we
disregard its explicit formula.

Identifying further: $\mathbf{L}\simeq\mathbb{A}^{1}$, via
$x\longmapsto x\sigma_{\mathbf{L}}$, and
$\mathbf{L}^{\prime}\simeq\mathbb{A}^{1}$, via $x\longmapsto
x\sigma_{\mathbf{L}^{\prime}}$, we get $\mathbf{V} \iso
\mathbb{A}^{1}\times\mathbb{A}^{1}$. In these coordinates the
Frobenius morphism is given by: $\Fr(x,y)=(y^{p},-x^{p})$. Now we
can write $\Fr_{\mathbf{L^{\circ}}}$ in coordinates:
\[
\Fr_{\mathbf{L^{\circ}}}(\mathcal{G)(}x)=
{\textstyle\int\limits_{y\in\mathbb{A}^{1}}}
\mathscr{L}_{\psi}(xy)\otimes\mathcal{G(}y^{p}).
\]

\subsubsection{Hecke eigensheaf}
We consider the sheaf of Deligne kernels $\mathcal{K}_{\rho}=\mathcal{K}%
_{\rho,\mathbf{L^{\circ}}}$ (Section \ref{Deligne's sheaf}).
Restricting to the torus $\mathbf{T},$ and using (\ref{iden1}) we
may consider $\mathcal{K}_{\rho}$ as a sheaf on $\mathbf{T}
\times\mathbf{L}^{\prime}\times\mathbf{L}^{\prime}$. It is a
direct computation to obtain the explicit formula:
$$\mathcal{K}_{\rho
}(g,x,y)=\mathscr{L}_{\chi_{_\mathrm{q}}}(g)\otimes\delta_{x-g^{-1}y},
$$
where $\mathscr{L}_{\chi_{_\mathrm{q}}}$ is the \textit{Kummer}
sheaf associated to the quadratic
character $\chi_{_\mathrm{q}}:\T\rightarrow \C^{\ast}$. The sheaf $\delta_{x-g^{-1}y}$ is obtained as $\delta_{x-g^{-1}%
y}=i_{!}\C _{\mathbf{L}^{\prime}}$, where
$i:\mathbf{L}^{\prime}\rightarrow
\mathbf{L}^{\prime}\times\mathbf{L}^{\prime}$,
$i(x,y)=(x,g^{-1}y)$.

Consider the imbedding
$j:\mathbf{T}\hookrightarrow\mathbf{L}^{\prime},\r
j(g)=g\sigma_{\mathbf{L}^{\prime}}$ and the sheaf:
$$\mathcal{S}_{\Psi_\chi}%
:=j_{!}(\mathscr{L}_{\chi_{_\mathrm{q}}}\otimes\mathscr{L}_\chi).$$
We have the following proposition:

\begin{proposition}\label{eigensheaf}
The sheaf $\mathcal{S}_{\Psi_\chi}$ is perverse , i.e.,
$\mathcal{S}_{\Psi_\chi}\in \Perv(\mathbf{L}^{\prime})$. Moreover,
\end{proposition}
\begin{enumerate}
\item It admits an $\mathbf{L^{\circ}}$-Weil structure, i.e., $\Fr_{\mathbf{L^{\circ
}}}(\mathcal{S}_{\Psi_\chi})\simeq\mathcal{S}_{\Psi_\chi}$.

\item It admits an $\mathscr{L}_\chi$-$\mathbf{T}$ equivariant
structure, i.e.,
$\mathcal{K}_{\rho}\ast\mathcal{S}_{\Psi_\chi}\simeq\mathscr{L}_\chi\boxtimes
\mathcal{S}_{\Psi_\chi}$.
\end{enumerate}

\subsubsection{Hecke eigenfunction}

Apply the Braverman-Polishchuk correspondence (\ref{BPC}) and
obtain the function:
$$
\Psi_\chi := [\mathcal{S}_{\Psi_\chi}]\in
\H_{\mathbf{L^{\circ}}}.$$ It is clear that this gives a
$\chi$-$\T$-eigenfunction.

\section{Proof of the Supremum Conjecture}

Let $\mathbf{L^{\circ}}=(\mathbf{L,}\sigma_{\mathbf{L}})$ be a
rational enhanced Lagrangian, i.e., $\Fr\,
\mathbf{L^{\circ}=L^{\circ}}$. Let $\L^{\circ
}=\mathbf{L^{\circ}}(\mathbb{F}_{p})$. Consider the associated
model $\H_{\mathbf{L^{\circ
}}}$. We have $\H_{\mathbf{L^{\circ}}}=\H_{\L^{\circ}}=\Gamma(\X_{\L^{\circ}%
},\mathcal{\H}_{\L^{\circ}})$, where $\mathcal{\H}_{\L^{\circ}}$
is an Hermitian line bundle on
$\X_{\L^{\circ}}=\widetilde{\L}\backslash \E$. Let $\Psi_{\chi
}^{\,\mathbf{L}^{\circ}}\in$ $\H_{\mathbf{L^{\circ}}}$ be a
$\chi$-$\T$-eigenvector. It will be convenient for us to take
$\Psi_\chi^{\,\mathbf{L}^{\circ}}$ with the normalization
$\left\Vert \Psi_\chi^{\,\mathbf{L}^{\circ}}\right\Vert ^{2}=q$.

Fix a point $x\in \X_{\L^{\circ}}$. Denote by $a_x$ the quantity:
$$
a_{x} :=\left\vert \Psi_{\chi
}^{\,\mathbf{L}^{\circ}}(x)\right\vert ^{2}.
$$
We would like to study the quantity $a_{x}$. In particular, we
will prove that independent of $p$, $\L^{\circ}$ and $\chi$ we
have:
\begin{proposition}[Supremum conjecture - restated]
\label{supremum} The following bound holds:
$$
a_{x}\leq 4,
$$
for every $x\in \X_{\L^{\circ}}$.
\end{proposition}

First we give the quantity $a_{x}$ a representation theoretic
interpretation. Consider the abelian subgroup $\L\subset \E$. It
acts semisimply on the space $\H_{\mathbf{L^{\circ}}}$, therefore
we obtain a decomposition:
\[
\H_{\mathbf{L^{\circ}}}=%
{\textstyle\bigoplus\limits_{\psi:\L\rightarrow \C^{\ast}}}
\H_{\psi},
\]
where $\psi$ runs over the set of characters of the group $\L$.
The point $x$ corresponds to a specific character
$\psi_{x}:\L\rightarrow \C^{\ast}$. Denote by
$P_{x}=P_{x}^{\mathbf{L^{\circ}}}=\frac{1}{\left\vert
L\right\vert }%
{\textstyle\sum\limits_{l\in \L}}
\psi_{x}(l)\pi_{_\mathbf{L^{\circ}}}(l)$ the orthogonal projector
on the space
$\H_{\psi_{x}}$. We have:%
\begin{equation*}
a_{x}=\left\langle P_{x}\Psi_\chi^{\,\mathbf{L}^{\circ}},\Psi_\chi^{\,\mathbf{L}%
^{\circ}}\right\rangle.
\end{equation*}

The main observation is, that the scalar $a_{x}$, which is defined
completely in representation theoretic terms, does not depend on
the specific model $\H_{\mathbf{L^{\circ}}}$. We are free to use a
different model for the computation. We choose a model in which
$\Psi_\chi$ has a convenient form. In more detail, let
$\mathbf{M^{\circ}}=(\mathbf{M,}\sigma_{\mathbf{M}})$ be such that
$\mathbf{M}$ is fixed by $\mathbf{\T}$. Let $\Psi_\chi
^{\,\mathbf{M^{\circ}}}=\SF_{\mathbf{M^{\circ},L^{\circ}}}(\Psi_{\chi
}^{\,\mathbf{L}^{\circ}})$. Then
$\Psi_\chi^{\,\mathbf{M^{\circ}}}\in \H_{\mathbf{M^{\circ}}}$ and
clearly $\Psi_\chi^{\,\mathbf{M^{\circ}}}$ is a $\chi
$-$\T$-eigenvector. We have:
\begin{equation}
a_{x}=\left\langle P_{x}\Psi_\chi^{\,\mathbf{M^{\circ}}},\Psi_\chi
^{\,\mathbf{M^{\circ}}}\right\rangle \label{scalar2},
\end{equation}

where now $P_{x}=P_{x}^{\mathbf{M^{\circ}}}=\frac{1}{\left\vert
L\right\vert}
{\textstyle\sum\limits_{l\in \L}}
\psi_{x}(l)\pi_{_\mathbf{M^{\circ}}}(l)$.
\\\\
Developing formula (\ref{scalar2}) we obtain:
$$
a_{x}=\frac{1}{\left\vert \L\right\vert }%
{\textstyle\sum\limits_{l\in \L}}
\psi_{x}(l)\left\langle \pi_{_\mathbf{M^{\circ}}}(l)\,\Psi_\chi%
^{\,\mathbf{M^{\circ}}},\Psi_\chi^{\,\mathbf{M^{\circ}}}\right\rangle.
$$

Next we study the function: $\left\langle
\pi_{_\mathbf{M^{\circ}}}(\cdot)\,\Psi_{\chi
}^{\,\mathbf{M^{\circ}}},\Psi_\chi^{\,\mathbf{M^{\circ}}}\right\rangle
:\E\rightarrow \C$. This is only a matrix coefficient. We will
analyze it using geometry.

Let $\mathcal{S}_{\Psi_\chi}^{\,\mathbf{M^{\circ}}}\in \mathcal{D}_{\mathbf{M^{\circ}%
}}^{\w}$ be an $\mathscr{L}_{\chi}$-$\mathbf{T}$-eigensheaf (see
Proposition \ref{eigensheaf}). We can assume that
$\Psi_\chi^{\,\mathbf{M^{\circ
}}}=[\mathcal{S}_{\Psi_\chi}^{\,\mathbf{M^{\circ}}}]$ (this is the
reason why we choose the specific normalization for $\Psi_\chi$).
We have $\left\langle
\pi_{_\mathbf{M^{\circ}}}(\cdot)\,\Psi_\chi^{,\mathbf{M^{\circ}}},\Psi_{\chi
}^{,\mathbf{M^{\circ}}}\right\rangle
=f^{\m(\mathcal{S}_{\Psi_\chi}^{\,\mathbf{M^{\circ
}}},\mathcal{S}_{\Psi_\chi}^{\,\mathbf{M^{\circ}}})}$. It is
enough to study the Weil sheaf
$\m(\mathcal{S}_{\Psi_\chi}^{\,\mathbf{M^{\circ}}},\mathcal{S}_{\Psi_\chi}
^{\,\mathbf{M^{\circ}}})\in \mathcal{D}_{\mathrm{w}}(\mathbf{H})$.
We can in fact compute
$\m(\mathcal{S}_{\Psi_\chi}^{\,\mathbf{M^{\circ}}},\mathcal{S}_{\Psi_\chi}%
^{\,\mathbf{M^{\circ}}})$ explicitly: Let
$\mathbf{M^{\circ\prime}=(M}^{\prime
},\sigma_{\mathbf{M}^{\prime}})=(\Fr
\,\mathbf{M,}\Fr\,\sigma_{\mathbf{M}})$. We can assume that
$\omega(\sigma_{\mathbf{M}},\sigma_{\mathbf{M}^{\prime}})=1$. We
have the decompositions: $\mathbf{V=M}^{\prime}\times\mathbf{M}$,
and $\mathbf{H=M}^{\prime}\times\mathbf{M\times Z}$. Recalling the
definition of a
matrix coefficient (\ref{matrix}), we obtain:%
\begin{align}
\mathrm{m}(\mathcal{S}_{\Psi_\chi}^{\,\mathbf{M^{\circ}}},\mathcal{S}_{\Psi_\chi}^{\,\mathbf{M^{\circ
}}})(m^{\prime},m,z)  & =%
{\textstyle\int\limits_{t\in\mathbf{M}^{\prime}}}
\mathscr{L}_{\psi}(z+%
\half
\omega(m^{\prime},m)+\omega(t,m))\otimes\label{formula1}\\
&
\otimes\mathcal{S}_{\Psi_\chi}^{\,\mathbf{M^{\circ}}}(t+m^{\prime})\otimes
\mathbb{D}(\mathcal{S}_{\Psi_\chi}^{\,\mathbf{M^{\circ}}})(t)\nonumber.
\end{align}

As was previously done, we can identify
$\mathcal{D}_{\mathbf{M^{\circ}},\psi
}^{\w}=\mathcal{D}(\mathbf{M}^{\prime})$ as plain triangulated
categories, but Weil
structure is realized by the functor $\Fr_{\mathbf{M^{\circ}}}:\mathcal{D}(\mathbf{M}%
^{\prime})\rightarrow \mathcal{D}(\mathbf{M}^{\prime})$. Taking
coordinates $\mathbf{M}^{\prime}\simeq\mathbb{A}^{1}$,
$x\longmapsto x\cdot\sigma _{\mathbf{M}^{\prime}}$, and
$\mathbf{M}\simeq\mathbb{A}^{1}$, $x\longmapsto
x\cdot\sigma_{\mathbf{M}}$, we can further identify
$\mathcal{D}(\mathbf{M}^{\prime })=\mathcal{D}(\mathbb{A}^{1})$ .
In these coordinates we have $\mathcal{S}_{\psi_\chi
}^{\mathbf{M^{\circ}}}=\mathscr{L}_{\chi}$. Writing formula
(\ref{formula1})
translates to:%
\[
\m(\mathcal{S}_{\Psi_\chi}^{\,\mathbf{M^{\circ}}},\mathcal{S}_{\Psi_\chi}^{\,\mathbf{M^{\circ
}}})(x^{\prime},x,z)=%
{\textstyle\int\limits_{t\in\mathbb{A}^{1}}}
\mathscr{L}_{\psi}(z-%
\half
x^{\prime}x+xt)\otimes\mathscr{L}_{\chi}(t+x)\otimes\mathscr{L}_{\chi^{-1}%
}(x^{\prime}).
\]

We have the following proposition:

\begin{proposition}
\label{matrixperv} The sheaf $\m(\mathcal{S}_{\Psi_\chi}^{\,\mathbf{M^{\circ}}%
},\mathcal{S}_{\Psi_\chi}^{\,\mathbf{M^{\circ}}})$ $\in
\Perv^{\w,1}(\mathbf{H})$, irreducible of pure weight 0, and it is
equivariant with respect to the action
of $\mathbf{T}_{\mathrm{A}}$. Moreover, $\m(\mathcal{S}_{\Psi_\chi}^{\,\mathbf{M^{\circ}}%
},\mathcal{S}_{\Psi_\chi}^{\,\mathbf{M^{\circ}}})$ is smooth on
the open set $U=\mathbf{H\backslash(M,}0)\cup(0,\mathbf{M)}$.
\end{proposition}

\bigskip

\textbf{Proof of proposition \ref{supremum}}. We have:
$$
{\textstyle\sum\limits_{l\in L}}
\psi_{x}(l)\left\langle \pi_{_\mathbf{M^{\circ}}}(l)\,\Psi_\chi%
^{\,\mathbf{M^{\circ}}},\Psi_\chi^{\,\mathbf{M^{\circ}}}\right\rangle =q+%
{\textstyle\sum\limits_{l\in \L^{\times}}}
\psi_{x}(l)\left\langle \pi_{_\mathbf{M^{\circ}}}(l)\,\Psi_\chi%
^{\,\mathbf{M^{\circ}}},\Psi_\chi^{\,\mathbf{M^{\circ}}}\right\rangle.
$$
It is enough to estimate $
{\textstyle\sum\limits_{l\in \L^{\times}}}
\psi_{x}(l)\left\langle \pi_{_\mathbf{M^{\circ}}}(l)\,\Psi_\chi%
^{\,\mathbf{M^{\circ}}},\Psi_\chi^{\,\mathbf{M^{\circ}}}\right\rangle
$. Consider the function:
$$
g=\psi_{x}(\cdot)\left\langle \pi_{_\mathbf{M^{\circ}}}%
(\cdot)\,\Psi_\chi^{\,\mathbf{M^{\circ}}},\Psi_\chi^{\,\mathbf{M^{\circ}}%
}\right\rangle :\L\rightarrow \C.$$ This function is obtained as
$g=f^{\mathcal{G}}$, where
$$
\SG := \SL_{\psi}\otimes
\m(\mathcal{S}_{\Psi_\chi}^{\,\mathbf{M^{\circ}}},\mathcal{S}_{\Psi_\chi}^{\,\mathbf{M^{\circ
}}})_{|\mathbf{L}^{\times}}.
$$
From proposition \ref{matrixperv} we deduce that
$\m(\mathcal{S}_{\Psi_\chi}^{\,\mathbf{M^{\circ}}},\mathcal{S}_{\Psi_\chi}%
^{\,\mathbf{M^{\circ}}})_{|\mathbf{L}^{\times}}$ is smooth,
therefore $\mathcal{G}$ is non-trivial and smooth. This implies
that the sheaf $\pi_{!}\mathcal{G}$ on the point, i.e.,
$\pi_{!}\mathcal{G}\in \mathcal{D}(pt)$, where $\pi$ denotes the
projection $\pi:\mathbf{L}^{\times}\rightarrow pt$, is supported
at degree 2, namely:
$$
\mathrm{H}^{i}(\pi_{!}\mathcal{G)=}0, \rev \text{for every}\,  i
\neq 2.
$$
Moreover, $\pi _{!}\mathcal{G}$ is of weight $\leq0$. This means
that the eigenvalues of Frobenius are bounded by:
$$
\left\vert \ev
\,(\Fr_{|\mathrm{H}^{2}(\pi_{!}\mathcal{G)}})\right\vert \leq q.
$$
Now, denoting by $d$ the dimension
$d:=\dim\,\mathrm{H}^{2}(\pi_{!}\mathcal{G)}$, we get:
$$
{\textstyle\sum\limits_{l\in \L^{\times}}}
\psi_{x}(l)\left\langle \pi_{_\mathbf{M^{\circ}}}(l)\,\Psi_\chi
^{\,\mathbf{M^{\circ}}},\Psi_\chi^{\,\mathbf{M^{\circ}}}\right\rangle
=f^{\pi _{!}\mathcal{G}}\leq dq.
$$
Hence,
\[
a_{x}\leq d+1.
\]

Therefore, by a direct computation, using the explicit formulas of
the sheaf $\mathcal{G}$, we show that $d = 3$. This completes the
proof. $\EProof$
\bigskip
\center\texttt{Israel, September 1, 2005.}
\end{document}

%% file: diagram.tex
\makeatletter

\def\diagram{\m@th\leftwidth=\z@ \rightwidth=\z@ \topheight=\z@
\botheight=\z@ \setbox\@picbox\hbox\bgroup}

\def\enddiagram{\egroup\wd\@picbox\rightwidth\unitlength
\ht\@picbox\topheight\unitlength \dp\@picbox\botheight\unitlength
\hskip\leftwidth\unitlength\box\@picbox}

\def\bfig{\begin{diagram}}
\def\efig{\end{diagram}}
\newcount\wideness \newcount\leftwidth \newcount\rightwidth
\newcount\highness \newcount\topheight \newcount\botheight

\def\ratchet#1#2{\ifnum#1<#2 \global #1=#2 \fi}

\def\putbox(#1,#2)#3{%
\horsize{\wideness}{#3} \divide\wideness by 2
{\advance\wideness by #1 \ratchet{\rightwidth}{\wideness}}
{\advance\wideness by -#1 \ratchet{\leftwidth}{\wideness}}
\vertsize{\highness}{#3} \divide\highness by 2
{\advance\highness by #2 \ratchet{\topheight}{\highness}}
{\advance\highness by -#2 \ratchet{\botheight}{\highness}}
\put(#1,#2){\makebox(0,0){$#3$}}}

\def\putlbox(#1,#2)#3{%
\horsize{\wideness}{#3}
{\advance\wideness by #1 \ratchet{\rightwidth}{\wideness}}
{\ratchet{\leftwidth}{-#1}}
\vertsize{\highness}{#3} \divide\highness by 2
{\advance\highness by #2 \ratchet{\topheight}{\highness}}
{\advance\highness by -#2 \ratchet{\botheight}{\highness}}
\put(#1,#2){\makebox(0,0)[l]{$#3$}}}

\def\putrbox(#1,#2)#3{%
\horsize{\wideness}{#3}
{\ratchet{\rightwidth}{#1}}
{\advance\wideness by -#1 \ratchet{\leftwidth}{\wideness}}
\vertsize{\highness}{#3} \divide\highness by 2
{\advance\highness by #2 \ratchet{\topheight}{\highness}}
{\advance\highness by -#2 \ratchet{\botheight}{\highness}}
\put(#1,#2){\makebox(0,0)[r]{$#3$}}}

\def\adjust[#1]{} 

\newcount \coefa
\newcount \coefb
\newcount \coefc
\newcount\tempcounta
\newcount\tempcountb
\newcount\tempcountc
\newcount\tempcountd
\newcount\xext
\newcount\yext
\newcount\xoff
\newcount\yoff
\newcount\gap%
\newcount\arrowtypea
\newcount\arrowtypeb
\newcount\arrowtypec
\newcount\arrowtyped
\newcount\arrowtypee
\newcount\height
\newcount\width
\newcount\xpos
\newcount\ypos
\newcount\run
\newcount\rise
\newcount\arrowlength
\newcount\halflength
\newcount\arrowtype
\newdimen\tempdimen
\newdimen\xlen
\newdimen\ylen
\newsavebox{\tempboxa}%
\newsavebox{\tempboxb}%
\newsavebox{\tempboxc}%

\newdimen\w@dth

\def\setw@dth#1#2{\setbox\z@\hbox{\m@th$#1$}\w@dth=\wd\z@
\setbox\@ne\hbox{\m@th$#2$}\ifnum\w@dth<\wd\@ne \w@dth=\wd\@ne \fi
\advance\w@dth by 1.2em}

\def\t@^#1_#2{\allowbreak\def\n@one{#1}\def\n@two{#2}\mathrel
{\setw@dth{#1}{#2}
\mathop{\hbox to \w@dth{\rightarrowfill}}\limits
\ifx\n@one\empty\else ^{\box\z@}\fi
\ifx\n@two\empty\else _{\box\@ne}\fi}}
\def\t@@^#1{\@ifnextchar_{\t@^{#1}}{\t@^{#1}_{}}}
\def\to{\@ifnextchar^{\t@@}{\t@@^{}}}

\def\t@left^#1_#2{\def\n@one{#1}\def\n@two{#2}\mathrel{\setw@dth{#1}{#2}
\mathop{\hbox to \w@dth{\leftarrowfill}}\limits
\ifx\n@one\empty\else ^{\box\z@}\fi
\ifx\n@two\empty\else _{\box\@ne}\fi}}
\def\t@@left^#1{\@ifnextchar_{\t@left^{#1}}{\t@left^{#1}_{}}}
\def\toleft{\@ifnextchar^{\t@@left}{\t@@left^{}}}

\def\two@^#1_#2{\allowbreak
\def\n@one{#1}\def\n@two{#2}\mathrel{\setw@dth{#1}{#2}
\mathop{\vcenter{\lineskip\z@\baselineskip\z@
                 \hbox to \w@dth{\rightarrowfill}%
                 \hbox to \w@dth{\rightarrowfill}}%
       }\limits
\ifx\n@one\empty\else ^{\box\z@}\fi
\ifx\n@two\empty\else _{\box\@ne}\fi}}
\def\tw@@^#1{\@ifnextchar _{\two@^{#1}}{\two@^{#1}_{}}}
\def\two{\@ifnextchar ^{\tw@@}{\tw@@^{}}}

\def\tofr@^#1_#2{\def\n@one{#1}\def\n@two{#2}\mathrel{\setw@dth{#1}{#2}
\mathop{\vcenter{\hbox to \w@dth{\rightarrowfill}\kern-1.7ex
                 \hbox to \w@dth{\leftarrowfill}}%
       }\limits
\ifx\n@one\empty\else ^{\box\z@}\fi
\ifx\n@two\empty\else _{\box\@ne}\fi}}
\def\t@fr@^#1{\@ifnextchar_ {\tofr@^{#1}}{\tofr@^{#1}_{}}}
\def\tofro{\@ifnextchar^ {\t@fr@}{\t@fr@^{}}}

\def\mon{\mathop{\m@th\hbox to
      14.6\P@{\lasyb\char'51\hskip-2.1\P@$\arrext$\hss
$\mathord\rightarrow$}}\limits} 
\def\leftmono{\mathrel{\m@th\hbox to
14.6\P@{$\mathord\leftarrow$\hss$\arrext$\hskip-2.1\P@\lasyb\char'50%
}}\limits} 
\mathchardef\arrext="0200       

\def\settypes(#1,#2,#3){\arrowtypea#1 \arrowtypeb#2 \arrowtypec#3}
\def\settoheight#1#2{\setbox\@tempboxa\hbox{#2}#1\ht\@tempboxa\relax}%
\def\settodepth#1#2{\setbox\@tempboxa\hbox{#2}#1\dp\@tempboxa\relax}%
\def\settokens`#1`#2`#3`#4`{%
     \def\tokena{#1}\def\tokenb{#2}\def\tokenc{#3}\def\tokend{#4}}
\def\setsqparms[#1`#2`#3`#4;#5`#6]{%
\arrowtypea #1
\arrowtypeb #2
\arrowtypec #3
\arrowtyped #4
\width #5
\height #6
}
\def\setpos(#1,#2){\xpos=#1 \ypos#2}

\def\settriparms[#1`#2`#3;#4]{\settripairparms[#1`#2`#3`1`1;#4]}%

\def\settripairparms[#1`#2`#3`#4`#5;#6]{%
\arrowtypea #1
\arrowtypeb #2
\arrowtypec #3
\arrowtyped #4
\arrowtypee #5
\width #6
\height #6
}

\def\resetparms{\settripairparms[1`1`1`1`1;500]\width 500}

\resetparms

\def\mvector(#1,#2)#3{
\put(0,0){\vector(#1,#2){#3}}%
\put(0,0){\vector(#1,#2){26}}%
}
\def\evector(#1,#2)#3{{
\arrowlength #3
\put(0,0){\vector(#1,#2){\arrowlength}}%
\advance \arrowlength by-30
\put(0,0){\vector(#1,#2){\arrowlength}}%
}}

\def\horsize#1#2{%
\settowidth{\tempdimen}{$#2$}%
#1=\tempdimen
\divide #1 by\unitlength
}

\def\vertsize#1#2{%
\settoheight{\tempdimen}{$#2$}%
#1=\tempdimen
\settodepth{\tempdimen}{$#2$}%
\advance #1 by\tempdimen
\divide #1 by\unitlength
}

\def\putvector(#1,#2)(#3,#4)#5#6{{%
\ifnum3<\arrowtype
\putdashvector(#1,#2)(#3,#4)#5\arrowtype
\else
\ifnum\arrowtype<-3
\putdashvector(#1,#2)(#3,#4)#5\arrowtype
\else
\xpos=#1
\ypos=#2
\run=#3
\rise=#4
\arrowlength=#5
\ifnum \arrowtype<0
    \ifnum \run=0
        \advance \ypos by-\arrowlength
    \else
        \tempcounta \arrowlength
        \multiply \tempcounta by\rise
        \divide \tempcounta by\run
        \ifnum\run>0
            \advance \xpos by\arrowlength
            \advance \ypos by\tempcounta
        \else
            \advance \xpos by-\arrowlength
            \advance \ypos by-\tempcounta
        \fi
    \fi
    \multiply \arrowtype by-1
    \multiply \rise by-1
    \multiply \run by-1
\fi
\ifcase \arrowtype
\or \put(\xpos,\ypos){\vector(\run,\rise){\arrowlength}}%
\or \put(\xpos,\ypos){\mvector(\run,\rise)\arrowlength}%
\or \put(\xpos,\ypos){\evector(\run,\rise){\arrowlength}}%
\fi\fi\fi
}}

\def\putsplitvector(#1,#2)#3#4{
\xpos #1
\ypos #2
\arrowtype #4
\halflength #3
\arrowlength #3
\gap 140
\advance \halflength by-\gap
\divide \halflength by2
\ifnum\arrowtype>0
   \ifcase \arrowtype
   \or \put(\xpos,\ypos){\line(0,-1){\halflength}}%
       \advance\ypos by-\halflength
       \advance\ypos by-\gap
       \put(\xpos,\ypos){\vector(0,-1){\halflength}}%
   \or \put(\xpos,\ypos){\line(0,-1)\halflength}%
       \put(\xpos,\ypos){\vector(0,-1)3}%
       \advance\ypos by-\halflength
       \advance\ypos by-\gap
       \put(\xpos,\ypos){\vector(0,-1){\halflength}}%
   \or \put(\xpos,\ypos){\line(0,-1)\halflength}%
       \advance\ypos by-\halflength
       \advance\ypos by-\gap
       \put(\xpos,\ypos){\evector(0,-1){\halflength}}%
   \fi
\else \arrowtype=-\arrowtype
   \ifcase\arrowtype
   \or \advance \ypos by-\arrowlength
       \put(\xpos,\ypos){\line(0,1){\halflength}}%
       \advance\ypos by\halflength
       \advance\ypos by\gap
       \put(\xpos,\ypos){\vector(0,1){\halflength}}%
   \or \advance \ypos by-\arrowlength
       \put(\xpos,\ypos){\line(0,1)\halflength}%
       \put(\xpos,\ypos){\vector(0,1)3}%
       \advance\ypos by\halflength
       \advance\ypos by\gap
       \put(\xpos,\ypos){\vector(0,1){\halflength}}%
   \or \advance \ypos by-\arrowlength
       \put(\xpos,\ypos){\line(0,1)\halflength}%
       \advance\ypos by\halflength
       \advance\ypos by\gap
       \put(\xpos,\ypos){\evector(0,1){\halflength}}%
   \fi
\fi
}

\def\putmorphism(#1)(#2,#3)[#4`#5`#6]#7#8#9{{%
\run #2
\rise #3
\ifnum\rise=0
  \puthmorphism(#1)[#4`#5`#6]{#7}{#8}#9%
\else\ifnum\run=0
  \putvmorphism(#1)[#4`#5`#6]{#7}{#8}#9%
\else
\setpos(#1)%
\arrowlength #7
\arrowtype #8
\ifnum\run=0
\else\ifnum\rise=0
\else
\ifnum\run>0
    \coefa=1
\else
   \coefa=-1
\fi
\ifnum\arrowtype>0
   \coefb=0
   \coefc=-1
\else
   \coefb=\coefa
   \coefc=1
   \arrowtype=-\arrowtype
\fi
\width=2
\multiply \width by\run
\divide \width by\rise
\ifnum \width<0  \width=-\width\fi
\advance\width by60
\if l#9 \width=-\width\fi
\putbox(\xpos,\ypos){#4}
{\multiply \coefa by\arrowlength
\advance\xpos by\coefa
\multiply \coefa by\rise
\divide \coefa by\run
\advance \ypos by\coefa
\putbox(\xpos,\ypos){#5} }%
{\multiply \coefa by\arrowlength
\divide \coefa by2
\advance \xpos by\coefa
\advance \xpos by\width
\multiply \coefa by\rise
\divide \coefa by\run
\advance \ypos by\coefa
\if l#9%
   \putrbox(\xpos,\ypos){#6}%
\else\if r#9%
   \putlbox(\xpos,\ypos){#6}%
\fi\fi }%
{\multiply \rise by-\coefc
\multiply \run by-\coefc
\multiply \coefb by\arrowlength
\advance \xpos by\coefb
\multiply \coefb by\rise
\divide \coefb by\run
\advance \ypos by\coefb
\multiply \coefc by70
\advance \ypos by\coefc
\multiply \coefc by\run
\divide \coefc by\rise
\advance \xpos by\coefc
\multiply \coefa by140
\multiply \coefa by\run
\divide \coefa by\rise
\advance \arrowlength by\coefa
\ifcase\arrowtype
\or \put(\xpos,\ypos){\vector(\run,\rise){\arrowlength}}%
\or \put(\xpos,\ypos){\mvector(\run,\rise){\arrowlength}}%
\or \put(\xpos,\ypos){\evector(\run,\rise){\arrowlength}}%
\fi}\fi\fi\fi\fi}}

\newcount\numbdashes \newcount\lengthdash \newcount\increment

\def\howmanydashes{
\numbdashes=\arrowlength \lengthdash=40
\divide\numbdashes by \lengthdash
\lengthdash=\arrowlength
\divide\lengthdash by \numbdashes
\increment=\lengthdash
\multiply\lengthdash by 3
\divide\lengthdash by 5
}

\def\putdashvector(#1)(#2,#3)#4#5{%
\ifnum#3=0 \putdashhvector(#1){#4}#5
\else
\ifnum#2=0
\putdashvvector(#1){#4}#5\fi\fi}

\def\putdashhvector(#1,#2)#3#4{{%
\arrowlength=#3 \howmanydashes
\multiput(#1,#2)(\increment,0){\numbdashes}%
{\vrule height .4pt width \lengthdash\unitlength}
\arrowtype=#4 \xpos=#1
\ifnum\arrowtype<0 \advance\arrowtype by 7 \fi
\ifcase\arrowtype
\or \advance\xpos by 10
    \put(\xpos,#2){\vector(-1,0){\lengthdash}}
    \advance\xpos by 40
    \put(\xpos,#2){\vector(-1,0){\lengthdash}}
\or \advance \xpos by 10
    \put(\xpos,#2){\vector(-1,0){\lengthdash}}
    \advance\xpos by  \arrowlength
    \advance\xpos by  -50
    \put(\xpos,#2){\vector(-1,0){\lengthdash}}
\or \advance\xpos by 10
    \put(\xpos,#2){\vector(-1,0){\lengthdash}}
\or \advance\xpos by \arrowlength
    \advance\xpos by -\lengthdash
    \put(\xpos,#2){\vector(1,0){\lengthdash}}
\or {\advance\xpos by 10
    \put(\xpos,#2){\vector(1,0){\lengthdash}}}
    \advance\xpos by \arrowlength
    \advance\xpos by -\lengthdash
    \put(\xpos,#2){\vector(1,0){\lengthdash}}
\or \advance\xpos by \arrowlength
    \advance\xpos by -\lengthdash
    \put(\xpos,#2){\vector(1,0){\lengthdash}}
    \advance\xpos by -40
    \put(\xpos,#2){\vector(1,0){\lengthdash}}
   \fi
}}

\def\putdashvvector(#1,#2)#3#4{{%
\arrowlength=#3 \howmanydashes
\ypos=#2 \advance\ypos by -\arrowlength
\multiput(#1,#2)(0,\increment){\numbdashes}%
    {\vrule width .4pt height \lengthdash\unitlength}
\arrowtype=#4 \ypos=#2
\ifnum\arrowtype<0 \advance\arrowtype by 7 \fi
\ifcase\arrowtype
\or \advance\ypos by \arrowlength \advance\ypos by -40
    \put(#1,\ypos){\vector(0,1){\lengthdash}}
    \advance\ypos by -40
    \put(#1,\ypos){\vector(0,1){\lengthdash}}
\or \advance\ypos by 10
    \put(#1,\ypos){\vector(0,1){\lengthdash}}
    \advance\ypos by \arrowlength \advance\ypos by -40
    \put(#1,\ypos){\vector(0,1){\lengthdash}}
\or \advance\ypos by \arrowlength \advance\ypos by -40
    \put(#1,\ypos){\vector(0,1){\lengthdash}}
\or \advance\ypos by 10
    \put(#1,\ypos){\vector(0,-1){\lengthdash}}
\or \advance\ypos by 10
    \put(#1,\ypos){\vector(0,-1){\lengthdash}}
    \advance\ypos by \arrowlength \advance\ypos by -40
    \put(#1,\ypos){\vector(0,-1){\lengthdash}}
\or \advance\ypos by 10
    \put(#1,\ypos){\vector(0,-1){\lengthdash}}
    \advance\ypos by 40
    \put(#1,\ypos){\vector(0,-1){\lengthdash}}
\fi
}}

\def\puthmorphism(#1,#2)[#3`#4`#5]#6#7#8{{%
\xpos #1
\ypos #2
\width #6
\arrowlength #6
\arrowtype=#7
\putbox(\xpos,\ypos){#3\vphantom{#4}}%
{\advance \xpos by\arrowlength
\putbox(\xpos,\ypos){\vphantom{#3}#4}}%
\horsize{\tempcounta}{#3}%
\horsize{\tempcountb}{#4}%
\divide \tempcounta by2
\divide \tempcountb by2
\advance \tempcounta by30
\advance \tempcountb by30
\advance \xpos by\tempcounta
\advance \arrowlength by-\tempcounta
\advance \arrowlength by-\tempcountb
\putvector(\xpos,\ypos)(1,0)\arrowlength\arrowtype
\divide \arrowlength by2
\advance \xpos by\arrowlength
\vertsize{\tempcounta}{#5}%
\divide\tempcounta by2
\advance \tempcounta by20
\if a#8 %
   \advance \ypos by\tempcounta
   \putbox(\xpos,\ypos){#5}%
\else
   \advance \ypos by-\tempcounta
   \putbox(\xpos,\ypos){#5}%
\fi}}

\def\putvmorphism(#1,#2)[#3`#4`#5]#6#7#8{{%
\xpos #1
\ypos #2
\arrowlength #6
\arrowtype #7
\settowidth{\xlen}{$#5$}%
\putbox(\xpos,\ypos){#3}%
{\advance \ypos by-\arrowlength
\putbox(\xpos,\ypos){#4}}%
{\advance\arrowlength by-140
\advance \ypos by-70
\ifdim\xlen>0pt
   \if m#8%
      \putsplitvector(\xpos,\ypos)\arrowlength\arrowtype
   \else
   \putvector(\xpos,\ypos)(0,-1)\arrowlength\arrowtype
   \fi
\else
   \putvector(\xpos,\ypos)(0,-1)\arrowlength\arrowtype
\fi}%
\ifdim\xlen>0pt
   \divide \arrowlength by2
   \advance\ypos by-\arrowlength
   \if l#8%
      \advance \xpos by-40
      \putrbox(\xpos,\ypos){#5}%
   \else\if r#8%
      \advance \xpos by40
      \putlbox(\xpos,\ypos){#5}%
   \else
      \putbox(\xpos,\ypos){#5}%
   \fi\fi
\fi
}}

\def\putsquarep<#1>(#2)[#3;#4`#5`#6`#7]{{%
\setsqparms[#1]%
\setpos(#2)%
\settokens`#3`%
\puthmorphism(\xpos,\ypos)[\tokenc`\tokend`{#7}]{\width}{\arrowtyped}b%
\advance\ypos by \height
\puthmorphism(\xpos,\ypos)[\tokena`\tokenb`{#4}]{\width}{\arrowtypea}a%
\putvmorphism(\xpos,\ypos)[``{#5}]{\height}{\arrowtypeb}l%
\advance\xpos by \width
\putvmorphism(\xpos,\ypos)[``{#6}]{\height}{\arrowtypec}r%
}}

\def\putsquare{\@ifnextchar <{\putsquarep}{\putsquarep%
   <\arrowtypea`\arrowtypeb`\arrowtypec`\arrowtyped;\width`\height>}}
\def\square{\@ifnextchar< {\squarep}{\squarep
   <\arrowtypea`\arrowtypeb`\arrowtypec`\arrowtyped;\width`\height>}}
\def\squarep<#1>[#2`#3`#4`#5;#6`#7`#8`#9]{{
\setsqparms[#1]
\diagram
\putsquarep<\arrowtypea`\arrowtypeb`\arrowtypec`
\arrowtyped;\width`\height>
(0,0)[#2`#3`#4`{#5};#6`#7`#8`{#9}]
\enddiagram
}}                                                 
\def\putptrianglep<#1>(#2,#3)[#4`#5`#6;#7`#8`#9]{{%
\settriparms[#1]%
\xpos=#2 \ypos=#3
\advance\ypos by \height
\puthmorphism(\xpos,\ypos)[#4`#5`{#7}]{\height}{\arrowtypea}a%
\putvmorphism(\xpos,\ypos)[`#6`{#8}]{\height}{\arrowtypeb}l%
\advance\xpos by\height
\putmorphism(\xpos,\ypos)(-1,-1)[``{#9}]{\height}{\arrowtypec}r%
}}

\def\putptriangle{\@ifnextchar <{\putptrianglep}{\putptrianglep
   <\arrowtypea`\arrowtypeb`\arrowtypec;\height>}}
\def\ptriangle{\@ifnextchar <{\ptrianglep}{\ptrianglep
   <\arrowtypea`\arrowtypeb`\arrowtypec;\height>}}
\def\ptrianglep<#1>[#2`#3`#4;#5`#6`#7]{{
\settriparms[#1]
\diagram
\putptrianglep<\arrowtypea`\arrowtypeb`
\arrowtypec;\height>
(0,0)[#2`#3`#4;#5`#6`{#7}]
\enddiagram
}}                                            

\def\putqtrianglep<#1>(#2,#3)[#4`#5`#6;#7`#8`#9]{{%
\settriparms[#1]%
\xpos=#2 \ypos=#3
\advance\ypos by\height
\puthmorphism(\xpos,\ypos)[#4`#5`{#7}]{\height}{\arrowtypea}a%
\putmorphism(\xpos,\ypos)(1,-1)[``{#8}]{\height}{\arrowtypeb}l%
\advance\xpos by\height
\putvmorphism(\xpos,\ypos)[`#6`{#9}]{\height}{\arrowtypec}r%
}}

\def\putqtriangle{\@ifnextchar <{\putqtrianglep}{\putqtrianglep
   <\arrowtypea`\arrowtypeb`\arrowtypec;\height>}}
\def\qtriangle{\@ifnextchar <{\qtrianglep}{\qtrianglep
   <\arrowtypea`\arrowtypeb`\arrowtypec;\height>}}
\def\qtrianglep<#1>[#2`#3`#4;#5`#6`#7]{{
\settriparms[#1]
\width=\height                                
\diagram
\putqtrianglep<\arrowtypea`\arrowtypeb`
\arrowtypec;\height>
(0,0)[#2`#3`#4;#5`#6`{#7}]
\enddiagram
}}

\def\putdtrianglep<#1>(#2,#3)[#4`#5`#6;#7`#8`#9]{{%
\settriparms[#1]%
\xpos=#2 \ypos=#3
\puthmorphism(\xpos,\ypos)[#5`#6`{#9}]{\height}{\arrowtypec}b%
\advance\xpos by \height \advance\ypos by\height
\putmorphism(\xpos,\ypos)(-1,-1)[``{#7}]{\height}{\arrowtypea}l%
\putvmorphism(\xpos,\ypos)[#4``{#8}]{\height}{\arrowtypeb}r%
}}

\def\putdtriangle{\@ifnextchar <{\putdtrianglep}{\putdtrianglep
   <\arrowtypea`\arrowtypeb`\arrowtypec;\height>}}
\def\dtriangle{\@ifnextchar <{\dtrianglep}{\dtrianglep
   <\arrowtypea`\arrowtypeb`\arrowtypec;\height>}}
\def\dtrianglep<#1>[#2`#3`#4;#5`#6`#7]{{
\settriparms[#1]
\width=\height                                
\diagram
\putdtrianglep<\arrowtypea`\arrowtypeb`
\arrowtypec;\height>
(0,0)[#2`#3`#4;#5`#6`{#7}]
\enddiagram
}}

\def\putbtrianglep<#1>(#2,#3)[#4`#5`#6;#7`#8`#9]{{%
\settriparms[#1]%
\xpos=#2 \ypos=#3
\puthmorphism(\xpos,\ypos)[#5`#6`{#9}]{\height}{\arrowtypec}b%
\advance\ypos by\height
\putmorphism(\xpos,\ypos)(1,-1)[``{#8}]{\height}{\arrowtypeb}r%
\putvmorphism(\xpos,\ypos)[#4``{#7}]{\height}{\arrowtypea}l%
}}

\def\putbtriangle{\@ifnextchar <{\putbtrianglep}{\putbtrianglep
   <\arrowtypea`\arrowtypeb`\arrowtypec;\height>}}
\def\btriangle{\@ifnextchar <{\btrianglep}{\btrianglep
   <\arrowtypea`\arrowtypeb`\arrowtypec;\height>}}
\def\btrianglep<#1>[#2`#3`#4;#5`#6`#7]{{
\settriparms[#1]
\width=\height                               
\diagram
\putbtrianglep<\arrowtypea`\arrowtypeb`
\arrowtypec;\height>
(0,0)[#2`#3`#4;#5`#6`{#7}]
\enddiagram
}}

\def\putAtrianglep<#1>(#2,#3)[#4`#5`#6;#7`#8`#9]{{%
\settriparms[#1]%
\xpos=#2 \ypos=#3
{\multiply \height by2
\puthmorphism(\xpos,\ypos)[#5`#6`{#9}]{\height}{\arrowtypec}b}%
\advance\xpos by\height \advance\ypos by\height
\putmorphism(\xpos,\ypos)(-1,-1)[#4``{#7}]{\height}{\arrowtypea}l%
\putmorphism(\xpos,\ypos)(1,-1)[``{#8}]{\height}{\arrowtypeb}r%
}}

\def\putAtriangle{\@ifnextchar <{\putAtrianglep}{\putAtrianglep
   <\arrowtypea`\arrowtypeb`\arrowtypec;\height>}}
\def\Atriangle{\@ifnextchar <{\Atrianglep}{\Atrianglep
   <\arrowtypea`\arrowtypeb`\arrowtypec;\height>}}
\def\Atrianglep<#1>[#2`#3`#4;#5`#6`#7]{{
\settriparms[#1]
\width=\height                                     
\diagram
\putAtrianglep<\arrowtypea`\arrowtypeb`
\arrowtypec;\height>
(0,0)[#2`#3`#4;#5`#6`{#7}]
\enddiagram
}}

\def\putAtrianglepairp<#1>(#2)[#3;#4`#5`#6`#7`#8]{{%
\settripairparms[#1]%
\setpos(#2)%
\settokens`#3`%
\puthmorphism(\xpos,\ypos)[\tokenb`\tokenc`{#7}]{\height}{\arrowtyped}b%
\advance\xpos by\height
\puthmorphism(\xpos,\ypos)[\phantom{\tokenc}`\tokend`{#8}]%
{\height}{\arrowtypee}b%
\advance\ypos by\height
\putmorphism(\xpos,\ypos)(-1,-1)[\tokena``{#4}]{\height}{\arrowtypea}l%
\putvmorphism(\xpos,\ypos)[``{#5}]{\height}{\arrowtypeb}m%
\putmorphism(\xpos,\ypos)(1,-1)[``{#6}]{\height}{\arrowtypec}r%
}}

\def\putAtrianglepair{\@ifnextchar <{\putAtrianglepairp}{\putAtrianglepairp%
   <\arrowtypea`\arrowtypeb`\arrowtypec`\arrowtyped`\arrowtypee;\height>}}
\def\Atrianglepair{\@ifnextchar <{\Atrianglepairp}{\Atrianglepairp%
   <\arrowtypea`\arrowtypeb`\arrowtypec`\arrowtyped`\arrowtypee;\height>}}

\def\Atrianglepairp<#1>[#2;#3`#4`#5`#6`#7]{{
\settripairparms[#1]
\settokens`#2`
\width=\height                                
\diagram
\putAtrianglepairp                            
<\arrowtypea`\arrowtypeb`\arrowtypec`
\arrowtyped`\arrowtypee;\height>
(0,0)[{#2};#3`#4`#5`#6`{#7}]
\enddiagram
}}

\def\putVtrianglep<#1>(#2,#3)[#4`#5`#6;#7`#8`#9]{{%
\settriparms[#1]%
\xpos=#2 \ypos=#3
\advance\ypos by\height
{\multiply\height by2
\puthmorphism(\xpos,\ypos)[#4`#5`{#7}]{\height}{\arrowtypea}a}%
\putmorphism(\xpos,\ypos)(1,-1)[`#6`{#8}]{\height}{\arrowtypeb}l%
\advance\xpos by\height
\advance\xpos by\height
\putmorphism(\xpos,\ypos)(-1,-1)[``{#9}]{\height}{\arrowtypec}r%
}}

\def\putVtriangle{\@ifnextchar <{\putVtrianglep}{\putVtrianglep
   <\arrowtypea`\arrowtypeb`\arrowtypec;\height>}}
\def\Vtriangle{\@ifnextchar <{\Vtrianglep}{\Vtrianglep
   <\arrowtypea`\arrowtypeb`\arrowtypec;\height>}}
\def\Vtrianglep<#1>[#2`#3`#4;#5`#6`#7]{{
\settriparms[#1]
\width=\height                                 
\diagram
\putVtrianglep<\arrowtypea`\arrowtypeb`
\arrowtypec;\height>
(0,0)[#2`#3`#4;#5`#6`{#7}]
\enddiagram
}}

\def\putVtrianglepairp<#1>(#2)[#3;#4`#5`#6`#7`#8]{{
\settripairparms[#1]%
\setpos(#2)%
\settokens`#3`%
\advance\ypos by\height
\putmorphism(\xpos,\ypos)(1,-1)[`\tokend`{#6}]{\height}{\arrowtypec}l%
\puthmorphism(\xpos,\ypos)[\tokena`\tokenb`{#4}]{\height}{\arrowtypea}a%
\advance\xpos by\height
\puthmorphism(\xpos,\ypos)[\phantom{\tokenb}`\tokenc`{#5}]%
{\height}{\arrowtypeb}a%
\putvmorphism(\xpos,\ypos)[``{#7}]{\height}{\arrowtyped}m%
\advance\xpos by\height
\putmorphism(\xpos,\ypos)(-1,-1)[``{#8}]{\height}{\arrowtypee}r%
}}

\def\putVtrianglepair{\@ifnextchar <{\putVtrianglepairp}{\putVtrianglepairp%
    <\arrowtypea`\arrowtypeb`\arrowtypec`\arrowtyped`\arrowtypee;\height>}}
\def\Vtrianglepair{\@ifnextchar <{\Vtrianglepairp}{\Vtrianglepairp%
    <\arrowtypea`\arrowtypeb`\arrowtypec`\arrowtyped`\arrowtypee;\height>}}
\def\Vtrianglepairp<#1>[#2;#3`#4`#5`#6`#7]{{
\settripairparms[#1]
\settokens`#2`
\diagram
\putVtrianglepairp                             
<\arrowtypea`\arrowtypeb`\arrowtypec`
\arrowtyped`\arrowtypee;\height>
(0,0)[{#2};#3`#4`#5`#6`{#7}]
\enddiagram
}}

\def\putCtrianglep<#1>(#2,#3)[#4`#5`#6;#7`#8`#9]{{%
\settriparms[#1]%
\xpos=#2 \ypos=#3
\advance\ypos by\height
\putmorphism(\xpos,\ypos)(1,-1)[``{#9}]{\height}{\arrowtypec}l%
\advance\xpos by\height
\advance\ypos by\height
\putmorphism(\xpos,\ypos)(-1,-1)[#4`#5`{#7}]{\height}{\arrowtypea}l%
{\multiply\height by 2
\putvmorphism(\xpos,\ypos)[`#6`{#8}]{\height}{\arrowtypeb}r}%
}}

\def\putCtriangle{\@ifnextchar <{\putCtrianglep}{\putCtrianglep
    <\arrowtypea`\arrowtypeb`\arrowtypec;\height>}}
\def\Ctriangle{\@ifnextchar <{\Ctrianglep}{\Ctrianglep
    <\arrowtypea`\arrowtypeb`\arrowtypec;\height>}}
\def\Ctrianglep<#1>[#2`#3`#4;#5`#6`#7]{{
\settriparms[#1]
\width=\height                               
\diagram
\putCtrianglep<\arrowtypea`\arrowtypeb`
\arrowtypec;\height>
(0,0)[#2`#3`#4;#5`#6`{#7}]
\enddiagram
}}                                           
\def\putDtrianglep<#1>(#2,#3)[#4`#5`#6;#7`#8`#9]{{%
\settriparms[#1]%
\xpos=#2 \ypos=#3
\advance\xpos by\height \advance\ypos by\height
\putmorphism(\xpos,\ypos)(-1,-1)[``{#9}]{\height}{\arrowtypec}r%
\advance\xpos by-\height \advance\ypos by\height
\putmorphism(\xpos,\ypos)(1,-1)[`#5`{#8}]{\height}{\arrowtypeb}r%
{\multiply\height by 2
\putvmorphism(\xpos,\ypos)[#4`#6`{#7}]{\height}{\arrowtypea}l}%
}}

\def\putDtriangle{\@ifnextchar <{\putDtrianglep}{\putDtrianglep
    <\arrowtypea`\arrowtypeb`\arrowtypec;\height>}}
\def\Dtriangle{\@ifnextchar <{\Dtrianglep}{\Dtrianglep
   <\arrowtypea`\arrowtypeb`\arrowtypec;\height>}}
\def\Dtrianglep<#1>[#2`#3`#4;#5`#6`#7]{{
\settriparms[#1]
\width=\height                              
\diagram
\putDtrianglep<\arrowtypea`\arrowtypeb`
\arrowtypec;\height>
(0,0)[#2`#3`#4;#5`#6`{#7}]
\enddiagram
}}                                          
\def\setrecparms[#1`#2]{\width=#1 \height=#2}%

\def\recursep<#1`#2>[#3;#4`#5`#6`#7`#8]{{\m@th
\width=#1 \height=#2
\settokens`#3`
\settowidth{\tempdimen}{$\tokena$}
\ifdim\tempdimen=0pt
  \savebox{\tempboxa}{\hbox{$\tokenb$}}%
  \savebox{\tempboxb}{\hbox{$\tokend$}}%
  \savebox{\tempboxc}{\hbox{$#6$}}%
\else
  \savebox{\tempboxa}{\hbox{$\hbox{$\tokena$}\times\hbox{$\tokenb$}$}}%
  \savebox{\tempboxb}{\hbox{$\hbox{$\tokena$}\times\hbox{$\tokend$}$}}%
  \savebox{\tempboxc}{\hbox{$\hbox{$\tokena$}\times\hbox{$#6$}$}}%
\fi
\ypos=\height
\divide\ypos by 2
\xpos=\ypos
\advance\xpos by \width
\bfig
\putCtrianglep<-1`1`1;\ypos>(0,0)[`\tokenc`;#5`#6`{#7}]%
\puthmorphism(\ypos,0)[\tokend`\usebox{\tempboxb}`{#8}]{\width}{-1}b%
\puthmorphism(\ypos,\height)[\tokenb`\usebox{\tempboxa}`{#4}]{\width}{-1}a%
\advance\ypos by \width
\putvmorphism(\ypos,\height)[``\usebox{\tempboxc}]{\height}1r%
\efig
}}

\def\recurse{\@ifnextchar <{\recursep}{\recursep<\width`\height>}}

\def\puttwohmorphisms(#1,#2)[#3`#4;#5`#6]#7#8#9{{%
\puthmorphism(#1,#2)[#3`#4`]{#7}0a
\ypos=#2
\advance\ypos by 20
\puthmorphism(#1,\ypos)[\phantom{#3}`\phantom{#4}`#5]{#7}{#8}a
\advance\ypos by -40
\puthmorphism(#1,\ypos)[\phantom{#3}`\phantom{#4}`#6]{#7}{#9}b
}}

\def\puttwovmorphisms(#1,#2)[#3`#4;#5`#6]#7#8#9{{%
\putvmorphism(#1,#2)[#3`#4`]{#7}0a
\xpos=#1
\advance\xpos by -20
\putvmorphism(\xpos,#2)[\phantom{#3}`\phantom{#4}`#5]{#7}{#8}l
\advance\xpos by 40
\putvmorphism(\xpos,#2)[\phantom{#3}`\phantom{#4}`#6]{#7}{#9}r
}}

\def\puthcoequalizer(#1)[#2`#3`#4;#5`#6`#7]#8#9{{%
\setpos(#1)%
\puttwohmorphisms(\xpos,\ypos)[#2`#3;#5`#6]{#8}11%
\advance\xpos by #8
\puthmorphism(\xpos,\ypos)[\phantom{#3}`#4`#7]{#8}1{#9}
}}

\def\putvcoequalizer(#1)[#2`#3`#4;#5`#6`#7]#8#9{{%
\setpos(#1)%
\puttwovmorphisms(\xpos,\ypos)[#2`#3;#5`#6]{#8}11%
\advance\ypos by -#8
\putvmorphism(\xpos,\ypos)[\phantom{#3}`#4`#7]{#8}1{#9}
}}

\def\putthreehmorphisms(#1)[#2`#3;#4`#5`#6]#7(#8)#9{{%
\setpos(#1) \settypes(#8)
\if a#9 %
     \vertsize{\tempcounta}{#5}%
     \vertsize{\tempcountb}{#6}%
     \ifnum \tempcounta<\tempcountb \tempcounta=\tempcountb \fi
\else
     \vertsize{\tempcounta}{#4}%
     \vertsize{\tempcountb}{#5}%
     \ifnum \tempcounta<\tempcountb \tempcounta=\tempcountb \fi
\fi
\advance \tempcounta by 60
\puthmorphism(\xpos,\ypos)[#2`#3`#5]{#7}{\arrowtypeb}{#9}
\advance\ypos by \tempcounta
\puthmorphism(\xpos,\ypos)[\phantom{#2}`\phantom{#3}`#4]{#7}{\arrowtypea}{#9}
\advance\ypos by -\tempcounta \advance\ypos by -\tempcounta
\puthmorphism(\xpos,\ypos)[\phantom{#2}`\phantom{#3}`#6]{#7}{\arrowtypec}{#9}
}}

\def\setarrowtoks[#1`#2`#3`#4`#5`#6]{%
\def\toka{#1}
\def\tokb{#2}
\def\tokc{#3}
\def\tokd{#4}
\def\toke{#5}
\def\tokf{#6}
}
\def\hex{\@ifnextchar <{\hexp}{\hexp<1000`400>}}
\def\hexp<#1`#2>[#3`#4`#5`#6`#7`#8;#9]{%
\setarrowtoks[#9]
\yext=#2 \advance \yext by #2
\xext=#1 \advance\xext by \yext
\bfig
\putCtriangle<-1`0`1;#2>(0,0)[`#5`;\tokb``\tokd]
\xext=#1 \yext=#2 \advance \yext by #2
\putsquare<1`0`0`1;\xext`\yext>(#2,0)[#3`#4`#7`#8;\toka```\tokf]
\advance \xext by #2
\putDtriangle<0`1`-1;#2>(\xext,0)[`#6`;`\tokc`\toke]
\efig
}
\makeatother